\documentclass[aip, amsmath, amssymb, graphicx, reprint]{revtex4-1}

\draft % marks overfull lines with a black rule on the right

\usepackage{graphicx}% Include figure files
\usepackage{dcolumn}% Align table columns on decimal point
\usepackage{bm}% bold math
\usepackage[utf8]{inputenc}
\usepackage[T1]{fontenc}
\usepackage{mathptmx}
\usepackage{amsmath}
\usepackage{xcolor}
\usepackage{url}            % simple URL typesetting
\usepackage{hyperref}       % hyperlinks

\DeclareMathOperator{\arctanh}{arctanh}
\DeclareMathOperator{\sign}{sign}
\DeclareMathOperator{\Tr}{Tr}

\begin{document}

%\preprint{}

\title{Effect of compression in molecular spin-crossover chains} %Title of paper



\author{A. \surname{Gudyma}}
\affiliation{Max Planck Institute of Microstructure Physics, Weinberg 2, 06120 Halle/Saale, Germany}

\author{Iu. \surname{Gudyma}}%
 \email{yugudyma@gmail.com}
\affiliation{Physical, Technical and Computer Sciences Institute of Yuriy Fedkovych Chernivtsi National University, 58012 Chernivtsi, Ukraine}%


\date{\today}

\begin{abstract}

In this work, we investigate thermodynamic properties of the one-dimensional (1D) spin-crossover molecular chain being a subject of a constant external pressure. 
Effective compressible degenerate Ising model is used as a theoretical framework.
Using transfer matrix formalism analytic results for the low spin -- high spin crossover were obtained.
We derive the exact expressions for the fraction of molecules in the high spin state, correlation function and heat capacity. 
We provide analysis of parameters region where the spin crossover takes place and demonstrate how pressure changes location of the crossover. 

\end{abstract}



\maketitle %\maketitle must follow title, authors, abstract and \pacs

\maketitle

\section{Introduction}
\label{sec:Introduction}

The research of bistable molecular systems is a challenging field of modern scientific study.
The magnetic spin transition associated with the spin crossover (SCO) phenomenon represents a paradigm of bistability at the molecular level that is of current interest because of potential applications in the development of new generations of electronic devices such as nonvolatil memories, molecular sensors and displays~\cite{jureschi2015pressure, gudyma2015kinetics, halcrow2013spin}.
The interconversion of two spin states is observed in iron(II) coordination compounds in octahedral surroundings.
In these ones the paramagnetic high spin state (HS, S = 2) can be switched reversibly to the low spin state (LS, S = 0) by several external stimuli such as temperature, pressure or light irradiation, yielding significant structural, magnetic, and optical changes~\cite{3d43d7-1, coronado2020molecular, 3d43d7-3, halcrow2013spin, 3d43d7-5}.
In general, the spin-crossover materials are the class of inorganic coordination complexes of the chemical elements with $3d^{4}$-$3d^{7}$ electronic configuration of the outer orbital which form the ligand environment with first-row transient metal ion centered in octahedral ligand field.
These complexes can be reversibly switched between spin states, resulting in different magnetic, structural or optical properties.

The microscopic Ising-like model can be used for describing the behavior of spin-crossover crystals at molecular level.
Different energies and degeneracies of the HS and LS states can be taken into account as an effective temperature dependent field.
Low dimensional iron(II) spin transitional materials were a subject of recent experimental studies in both 1D~\cite{Experiment1d-1, sugahara2017control, nebbali2018one, wolny2020vibrational, Experiment1d-Bronisz, Garcia2015Two-Step} and 2D~\cite{Experiment2d-1, Experiment2d-2, Experiment2d-3} with various techniques and setups.
Note that the finite-size effects are important for understanding of the practical application of real low dimensional system.
In one dimension such materials may be described by Ising-like models and many important results obtained analytically~\cite{Spin-Crossover1d-1, Spin-Crossover1d-2, Spin-Crossover1d-3, PhysRevE.99.042117, HUTAK2021127020, gudyma20211d}. 
The one-dimensional (1D) Ising-like model plays an important role in statistical physics, being one of the models which have been solved exactly.
Compressible Ising model also has long history of study~\cite{zagrebnov1972spin, Salinas_1973, Henriques_1987}, and new results were obtained recently by numeric techniques~\cite{Numeric-1, Numeric-2, Numeric-3, Numeric-4, Numeric-5, Numeric-6}.
Real quasi-1D spin-crossover materials almost perfectly correspond to the one-dimensional Ising model causing particular interest for theoretical studies.

Elastic degrees of freedom cause change of the thermodynamic properties of the system. 
It is known that the free HS ferrous ion has the larger volume than the LS one.
Due to the difference in effective volume of HS and LS chains of spin crossover materials are sensitive to external pressure~\cite{Pressure-1}.
Therefore, the pressure becomes an important parameter for describing the system.
For example, the influence of pressure has been used to tune the spin transition properties of such 1D chain compounds.
In previous papers~\cite{gudyma2014diffusionless, gudyma2014study, gudyma2016spin} by one of authors, the deformations were considered as homogeneous and isotropic.  
Such compressible model  is the simplest  special  case of consideration of elastic nature of molecular crystals.
In this work we study effects of the constant external pressure on the thermodynamics properties of the spin crossover materials.

The outline of this work is as follows.
Sec.~\ref{sec:Model} defines the model's formalism.
In Sec.~\ref{sec:Partition function} we calculate the partition function and introduce effective Ising-like Hamiltonian with temperature dependent ferromagnetic constant and magnetic field. 
Given model is solved analytically using transfer matrix formalism, which we introduce in the Sec.~\ref{sec:Transfer_matrix_formalism}.
We demonstrate on the example of the system's volume and the correlation function how to make exact finite $N$ calculations in Sec.~\ref{sec:Volume}.
The specific heat capacity and susceptibility are obtained in Sec.~\ref{sec:Heat_capacity}.
In the remaining part of the manuscript, we focus on analytical and numerical results for spin-crossover molecular chain under the pressure.
Finally, results and discussions are given in Sec.~\ref{sec:Summary}.

\section{Model}
\label{sec:Model}

\begin{figure*}[!htb]
    \centering
    \includegraphics[width=0.4\columnwidth]{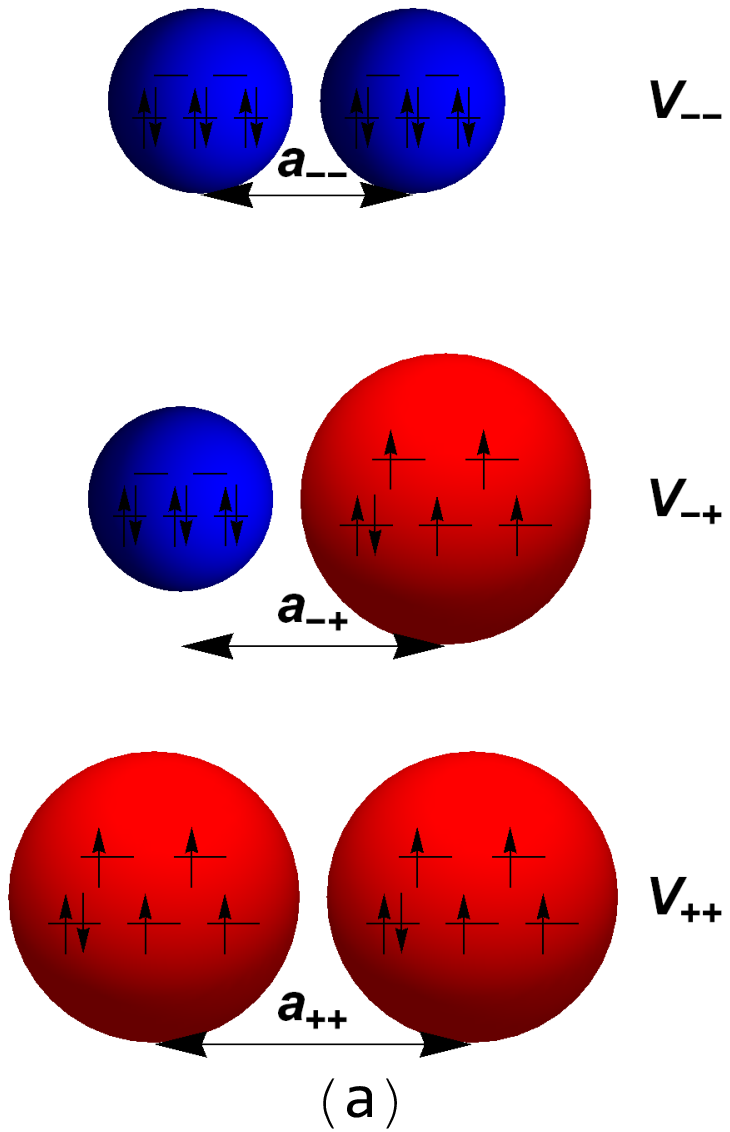}
    \includegraphics[width=0.99\columnwidth]{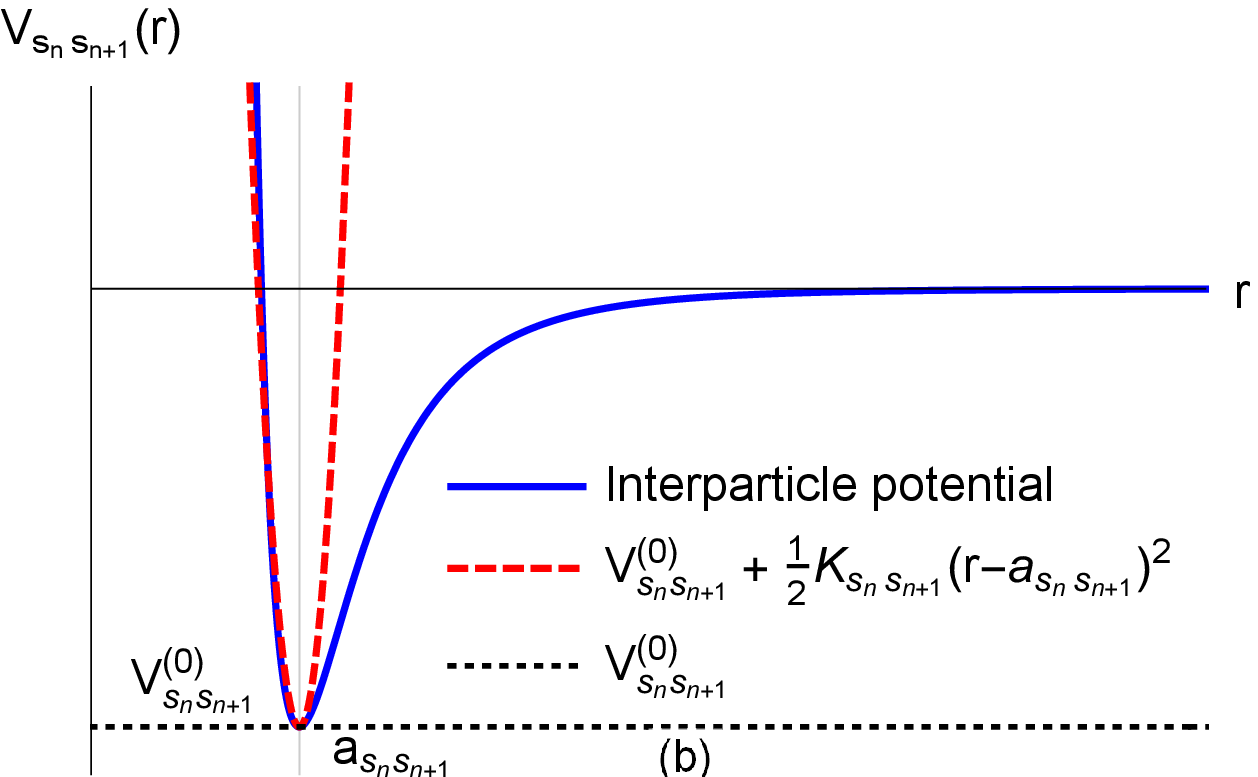}
    \caption{Schematic interactions of the pseudo-spin states and treatment of the inter-particle potential. (a) All possible configurations of the nearest pseudo-spin states. Interaction potentials and average distances between particles depend on the pseudo-spin states. (b) Interaction potential and harmonic approximation. We consider possible displacement of the particles from the equilibrium position for the given pseudo-state configuration to be small.}
    \label{fig:pseudospins_interactions}
\end{figure*}

In this work, we study behavior of a molecular chain under the external pressure.
Each particle in the chain may be in one of two states which have different properties, and may freely switch from one state to another.
We denote these states as the high spin (HS) pseudo-state and low spin (LS) pseudo-state. 
We introduce single-particle quasi-spin operator $\hat{s}$ as an operator which has eigenvalue $+1$ for the HS state and eigenvalue $-1$ for the LS state.
Let's denote the degeneracy of the pseudo-spin states $s_n$ as $g_{s_n}$, where $g_{s_n}=g_+$ for spin $s_n = +1$ pseudo-state and $g_{s_n} = g_-$ for  $s_n = -1$ pseudo-state.
We assume pair interactions of the molecules in the LS-LS, LS-HS and HS-HS pairs are different and we denote the corresponding pair potentials as $V_{--}(r)$, $V_{+-}(r)$ and $V_{++}(r)$.
In Fig.~\ref{fig:pseudospins_interactions}a we schematically illustrate all possible pseudo-spin configurations of the pairs of molecules.
These potentials directly correspond to the LS, HL and HS elastic potentials of the two-variable anharmonic Ising-like model~\cite{Elastic-potential-1, Spin-Crossover1d-2}.
Specific parameters of the interaction potentials can be extracted from experimental measurements, like X-ray diffraction~\cite{Experiment1d-Lecomte} or Brillouin spectroscopy~\cite{Experiment1d-Gutlich}.
The Hamiltonian of system consists of the Hamiltonian of the molecular chain and term describing action of the external pressure
\begin{equation}
    \hat{H} = \hat{H}_{MC} + \hat{H}_P.
    \label{eq:total_Hamiltonian}
\end{equation}
The molecular chain Hamiltonian is a sum of the pair potentials and single particle field
\begin{equation}
    \hat{H}_{MC} = \sum_{n=1}^{N-1} V_{s_ns_{n+1}}(x_n-x_{n+1}) + \sum_{n=1}^{N} W_{s_n},
    \label{eq:Initial_Hamiltonian}
\end{equation}
where $N$ is the total number of molecules in the chain and $W_{s_n}$ is the energy of the single-molecule pseudo-state.
The difference of the pseudo-state energies $\Delta = W_{+} - W_{-}$ is the external ligand field acting on a single molecule.
Action of the external pressure $P$ is described by the following extra term in the Hamiltonian
\begin{equation}
    \hat{H}_p = P L ,
\end{equation}
where $L = x_N - x_1$ is the effective volume of the one-dimensional system.
We apply an harmonic approximation for the nearest-neighbor pair potential $V_{s_ns_{n+1}}(r)$ at the potential minimum 
\begin{equation}
    V_{s_ns_{n+1}}(r) = V_{s_ns_{n+1}}^{(0)} + \frac{1}{2} K_{s_ns_{n+1}} \left(r-a_{s_ns_{n+1}}\right)^2,
\end{equation}
where $a_{s_ns_{n+1}}$ is the average distance between the particles at the equilibrium, $V_{s_ns_{n+1}}^{(0)} = V_{s_ns_{n+1}}(a_{s_ns_{n+1}})$ is the potential depth and $K_{s_ns_{n+1}}$ is an elastic constant coupling $n$-th and $(n+1)$-st molecules in the pseudo-states $s_n$ and $s_{n+1}$ respectively.
In Fig.~\ref{fig:pseudospins_interactions}b we illustrated treatment of the $V_{--}(r)$, $V_{+-}(r)$ and $V_{++}(r)$ potentials in the harmonic approximation.
Let's introduce relative coordinate variables $q_n = x_n-x_{n+1}$.
In the new variables $\hat{H}_P = \sum_{n=1}^{N-1} P q_n$.
We split total Hamiltonian (\ref{eq:total_Hamiltonian}) into a sum of two terms 
\begin{equation}
    \hat{H}_{MC} = \hat{H}_1 + \hat{H}_2,
    \label{eq:Separated_Hamiltonian}
\end{equation}
where
\begin{equation}
    \hat{H}_1 = \sum_{n=1}^{N-1} V_{s_ns_{n+1}}^{(0)}  + \sum_{n=1}^{N} W_{s_n},
    \label{eq:Separated_Hamiltonian_1}
\end{equation}
and
\begin{equation}
    \hat{H}_2 = \frac{1}{2} \sum_{n=1}^{N-1} K_{s_ns_{n+1}} \left(q_n-a_{s_ns_{n+1}}\right)^2 +  \sum_{n=1}^{N-1} P q_n .
\end{equation}
After some manipulations the Hamiltonian~$\hat{H}_1$ yields the Ising-like form~\cite{gudyma20211d}, and effects of degeneracy and the Hamiltonian $\hat{H}_2$ can be considered as the pressure and temperature dependent corrections to the coefficients of the basic Ising model.

\section{Partition function and effective Hamiltonian}
\label{sec:Partition function}

The partition function completely determines the statistical properties of the model. 
By the definition
\begin{equation}
    Z\! = \! \sum_{\langle s_1, \ldots, s_N \rangle}\! \iiint dq_1 \cdots dq_{N-1} g_{s_1} \cdots g_{s_N} e^{-\beta E\left( q_1, \ldots, q_{N-1}, s_1, \ldots, s_N \right)}
    %\\ = \sum_{\langle s_1, \ldots, s_N \rangle} g_{s_1} \cdots g_{s_N} \left( \prod_{n=1}^{N-1} \sqrt{\frac{2 \pi}{\beta K_{s_ns_{n+1}}}}  \right)  e^{-\beta E_1} 
     ,
    \label{eq:partition_function}
\end{equation}
where $E\left(q_1, \ldots, q_{N-1}, s_1, \ldots, s_N \right)$ is the energy, $\beta = 1/(k_{B}T)$ is the inverse temperature, $k_{B}$ denotes the Boltzmann constant and the sum goes over all possible spin configurations $\langle s_1, \ldots, s_N \rangle$.
Integration over phonon variables $q_n$ gives the expression
\begin{equation}
    Z = \sum_{\langle s_1, \ldots, s_N \rangle} g_{s_1} \cdots g_{s_N} 
    \left( \prod_{n=1}^{N-1} 
     \sqrt{\frac{2 \pi}{\beta K_{s_ns_{n+1}}}} e^{\frac{\beta P^2}{2 K_{s_ns_{n+1}} } - \beta P a_{s_ns_{n+1}}} \right)  e^{-\beta E_1}.
     \label{eq:partition_function_rewritten}
\end{equation}

We rewrite first part of the Hamiltonian (\ref{eq:Separated_Hamiltonian_1}) in terms of pseudo-spin variables
\begin{equation}
    \hat{H}_1 =  E_0 - \sum_{n=1}^{N-1} J s_n s_{n+1} - \sum_{n=1}^{N-1} B \frac{s_n+s_{n+1}}{2} - W(s_1) - W(s_N),
\end{equation}
where following notations were introduced $E_0 = \frac{N-1}{4} \left( V_{--}^{(0)} + V_{++}^{(0)} \right) + \frac{N-1}{2} V_{+-}^{(0)} + N \frac{W_{+} + W_{-}}{2}$,
 $J = - \frac{1}{4} \left( V_{--}^{(0)} + V_{++}^{(0)} \right) + \frac{1}{2} V_{+-}^{(0)}$,
 $B = \frac{1}{4} \left( V_{++}^{(0)} - V_{--}^{(0)} \right) - \frac{\Delta}{2}$,
and the term acting on the edge spins $W(s_n) = - \frac{\Delta}{4} s_n$.

We express spin state degeneracies as follows
\begin{equation}
    g_{s_n} %= e^{\ln g_{s_n}} 
    = e^{ \frac{1}{2} \left( \ln  g_{+} + \ln  g_{-} \right) + \frac{1}{2} \left( \ln  g_{+} - \ln  g_{-} \right) s_n}.
\end{equation}
The expression in the partition function~(\ref{eq:partition_function_rewritten}) which we obtain during the integration over the phononic degrees of freedom we rewrite in the form
\begin{multline}
    \sqrt{\frac{2 \pi}{\beta K_{s_ns_{n+1}}}}  e^{\frac{\beta P^2}{2 K_{s_ns_{n+1}} } - \beta P a_{s_ns_{n+1}}}
    \\ = e^{ \epsilon_{P} + \delta \epsilon + ( b_P + \delta b) (s_n+s_{n+1})/2 + ( j_P + \delta j ) s_n s_{n+1}} .
\end{multline}
where the energy term $\epsilon_P = - \beta  P a_\epsilon + \frac{\beta  P^2}{2 K_\epsilon}$, and the coefficients $j_P = - \beta  P a_J + \frac{\beta  P^2}{2 K_J }$ and $b_P = - \beta  P a_b + \frac{\beta  P^2}{2 K_B }$, 
with $a_\epsilon = \frac{1}{4} \left( a_{--} + a_{++} \right) + \frac{1}{2} a_{+-}$, $a_J = \frac{1}{4} \left( a_{--} + a_{++} \right) - \frac{1}{2} a_{+-}$ and $a_B = \frac{1}{2} \left( a_{++} - a_{--} \right)$, 
$\frac{1}{K_\epsilon} = \frac{1}{4} \left( \frac{1}{K_{--}} + \frac{1}{K_{++}} \right) + \frac{1}{2} \frac{1}{K_{+-}}$, $\frac{1}{K_J} = \frac{1}{4} \left( \frac{1}{K_{--}} + \frac{1}{K_{++}} \right) - \frac{1}{2} \frac{1}{K_{+-}}$ and $\frac{1}{K_B} = \frac{1}{2} \left( \frac{1}{K_{++}} - \frac{1}{K_{--}} \right)$, originates from the presence of pressure. 
The term $\delta \epsilon = - \frac{1}{8} \ln \left( \frac{\beta^4}{(2 \pi)^4} K_{+-}^2 K_{--} K_{++}  \right)$, the coefficients $\delta j = \frac{1}{8} \ln \left(  \frac{K_{+-} ^2}{ K_{--} K_{++} }  \right)$, and $\delta b = \frac{1}{4} \ln \left( \frac{K_{--}}{ K_{++} } \right)$ are manifestations of the elastic interaction.

Hence we have an expression for the partition function
\begin{equation}
    Z = \sum_{\langle s_1, \ldots, s_N \rangle}  e^{\epsilon +\sum_{n=1}^{N-1} v(s_n,s_{n+1}) + w(s_1) + w(s_N)  },
    \label{eq:partition_function_free_boundary}
\end{equation}
where 
\begin{equation}
    w(s_{n}) = w \frac{ s_{n}}{2},
\end{equation}
with the field acting on the edges $w = \frac{1}{2} \ln g - \frac{ \beta \Delta }{2}$, and effective two-particle energy terms 
\begin{equation}
    v(s_n,s_{n+1}) =  j s_n s_{n+1} + b (s_n + s_{n+1})/2,
\end{equation}
and 
\begin{subequations}
\begin{equation}
    \epsilon = \epsilon_P - \beta E_0 + \frac{N}{2}  \ln \left(  g_{+} g_{-} \right) + (N-1) \delta \epsilon,
\end{equation}
\begin{equation}
    j  = j_P + \beta J + \delta j,
\end{equation}
\begin{equation}
    b  = b_P + \beta B + \frac{1}{2} \ln g + \delta b.
\end{equation}
\label{eq:effective_parameters}
\end{subequations}
We make notation $g = \frac{g_{+}}{g_{-}}$.

The partition function (\ref{eq:partition_function_free_boundary}) can be expressed as the partition function of the Ising-like model with the effective Hamiltonian
\begin{multline}
    \hat{H}_{eff} = E_{0,eff} - \sum_{n=1}^{N-1} J_{eff}  \hat{s}_n \hat{s}_{n+1}  - \sum_{n=2}^{N-1} B_{eff}  \hat{s}_n  \\ + \frac{B_{boundary}+B_{eff}}{2} (\hat{s}_1 +\hat{s}_N) , 
    \label{eq:H_eff}
\end{multline}
where the reference energy $E_{0,eff} = E_0 - \frac{N k_B T}{2} \ln  g_{+}  g_{-}  - (N-1) \delta \epsilon k_B T -  (N-1) P a_\epsilon + (N-1) \frac{ P^2}{2 K_\epsilon}$, the ferromagnetic interaction constant $J_{eff} = J + \delta j k_B T +  P a_J - \frac{  P^2}{2 K_J }$, the field acting on the bulk $B_{\text{eff}} = B + \frac{k_B T}{2} \ln g + \delta b k_B T +  P a_b - \frac{ P^2}{2 K_B }$ and field acting on the boundaries $B_{boundary} = -\frac{\Delta}{2} +\frac{k_B T}{2} \ln g$. 
The effective Hamiltonian coincides with the Hamiltonian of the Ising model in which the reference energy, effective magnetic field~\cite{bousseksou1992ising, boukheddaden2000one} and ferromagnetic interaction constant are functions of temperature and pressure.
This dependence on temperature roots from the taking into account pseudo-states degeneracy and phononic interactions.

\section{Transfer-matrix formalism}
\label{sec:Transfer_matrix_formalism}

Thermodynamic properties of the system are completely described by the partition function.
Here, we use the transfer matrix formalism~\cite{Linares2004, Boukheddaden2007, gudyma20211d} to calculate the partition function.  
We rewrite the partition function (\ref{eq:partition_function_free_boundary}) in the following form
\begin{equation}
    Z = e^\epsilon \Tr \hat{T}^{N-1} \hat{R} ,
    \label{eq:partition_function_non_periodic}
\end{equation}
where the transfer matrix is
\begin{equation}
    \hat{T} = e^{v(s_n,s_{n+1})} =
    \begin{pmatrix}
        e^{ j + b } 
        & e^{ -j  } \\
        e^{ -j  } 
        & e^{ j - b } 
    \end{pmatrix},
\end{equation}
and the matrix $\hat{R}$ is accounting effects of the field acting on the surface spins
\begin{equation}
    \hat{R} = e^{ w(s_N) + w(s_1) } 
    =  \begin{pmatrix}
        e^{  w  } 
        & 1 \\
        1
        & e^{ - w  } 
    \end{pmatrix}.
\end{equation}

\begin{figure*}[!htb]
    \centering
    \includegraphics[width=0.9\columnwidth]{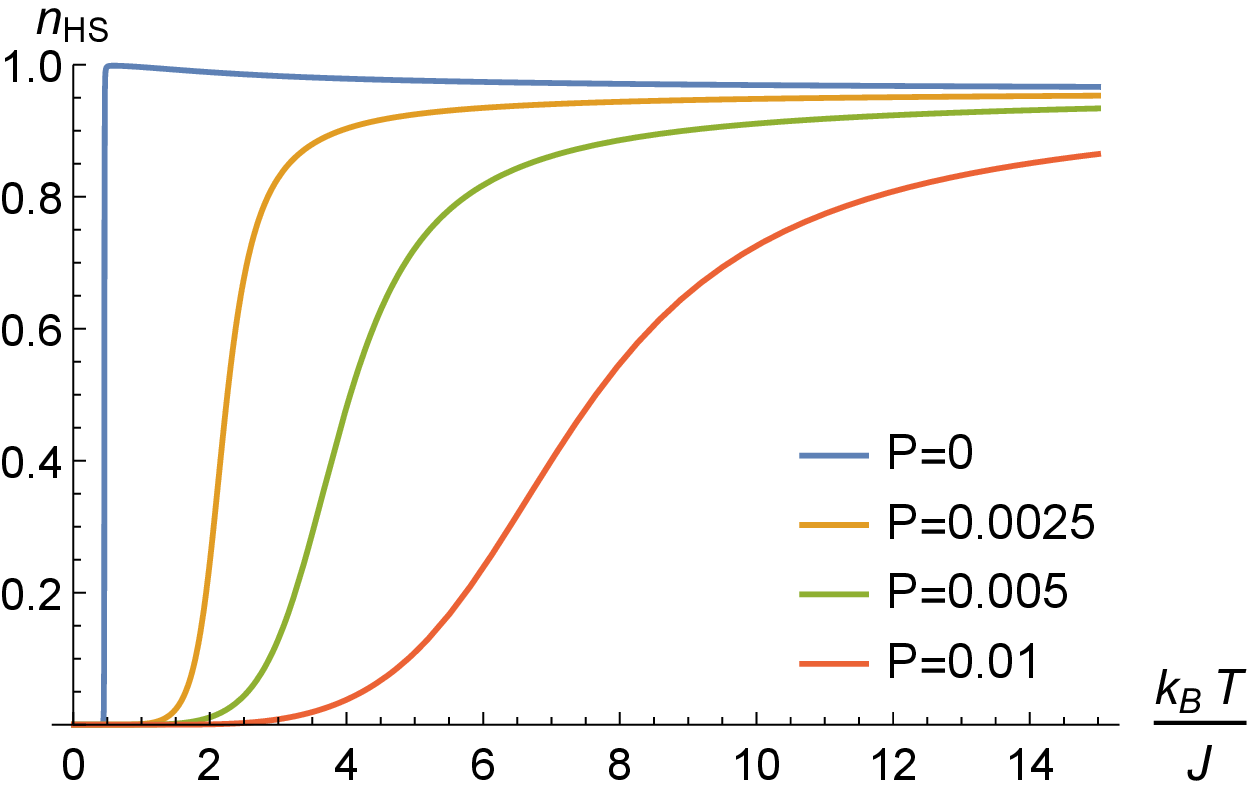}
    \includegraphics[width=0.9\columnwidth]{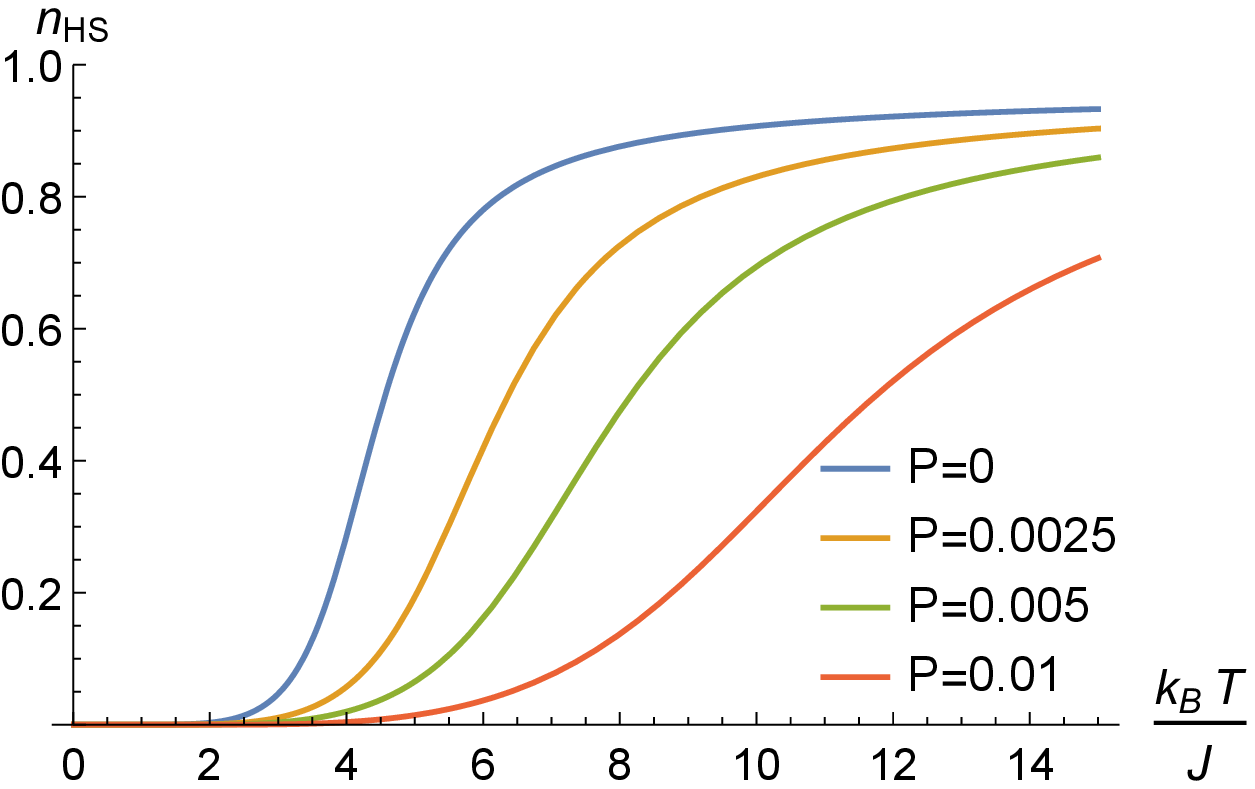}
    \caption{(a)-(b) Average occupation number $n_{HS}(T)$ as a function of temperature for $T_{eq}<T_{crossover}$ and $T_{eq}>T_{crossover}$ and various values of pressure $P=0, 0.0025, 0.005, 0.01$, and $\frac{1}{2}\ln g + \delta b =  0.5$ and $\delta j = 0.35$.
    %In the case $T_{eq}<T_{crossover}$ average magnetization has maximum at $T = T_0$.
    %For $T_{eq}=T_{crossover}$, maximum is reached at $T_0 = \infty$.
    %When $T_{eq}>T_{crossover}$ average magnetization is a monotonous function of temperature and has no extrema.
    }
    \label{fig:Average_magnetization_plots}
\end{figure*}

%Eigenvalues of the transfer matrix are given by the Eq.~(\ref{eq:lambda+}). 
For calculating $\Tr \hat{T}^{N-1} \hat{R}$ we change the basis to the eigenbasis of the transfer matrix  
\begin{equation}
    Z = e^\epsilon \Tr \hat{U} \hat{U}^{-1} \hat{T}^{N-1} \hat{U} \hat{U}^{-1} \hat{R},
    \label{eq:partition_function_non_periodic_1}
\end{equation}
where 
\begin{equation}
    \hat{U} = 
        \begin{pmatrix}
        \cos \phi
        & \sin \phi \\
        - \sin \phi
        & \cos \phi
    \end{pmatrix},
\end{equation}
and the angle of rotation $\phi$ is given by a solution of the equation
\begin{equation}
    \cot 2 \phi = e^{2 j } \sinh (b ).
\end{equation}
The eigenvalues of the transfer matrix $\hat{T}$ are
\begin{equation}
    \lambda_{\pm} = \left( e^{j } \cosh b  \pm \sqrt{e^{2 j } \sinh^2 b  + e^{-2 j }} \right) .
    \label{eq:lambda+}
\end{equation}
Therefore we obtain the partition function for the system of $N$ particles
\begin{equation}
    Z = e^\epsilon \left( c_{+} \lambda_{+}^{N-1} + c_{-} \lambda_{-}^{N-1} \right),
    \label{eq:partition_function_final}
\end{equation}
where the coefficients are
\begin{subequations}
\begin{equation} 
    c_{+} = \cosh w + \frac{e^{-2j} + \sinh b \sinh w }{\sqrt{\sinh^2 b  + e^{-4 j }}}, 
    \label{eq:c+}
\end{equation}
\begin{equation}
    c_{-} = \cosh w - \frac{e^{-2j} + \sinh b \sinh w }{\sqrt{\sinh^2 b  + e^{-4 j }}}.
\end{equation}
\end{subequations}
Until now all calculation were exact and the partition function (\ref{eq:partition_function_final}) contains all finite $N$ effects.
The free energy density is given by the following expression
\begin{equation}
    f = -\frac{1}{N \beta} \ln Z .
\end{equation}
In the thermodynamic limit, we obtain
\begin{equation}
    f = - \lim_{N \to \infty} \frac{1}{\beta N} \ln Z  = - \frac{\epsilon}{N \beta} - \frac{1}{\beta} \ln \lambda_+.
\end{equation}
Average magnetization per quasi-spin is $m = \left\langle s \right\rangle = \frac{1}{N} \frac{\partial \ln Z}{\partial b }$.
The magnetization per spin in the thermodynamic limit $\left\langle s \right\rangle$ at nonzero temperature $T$, pressure $P$, and external field $B$ is easily evaluated: 
\begin{equation}
    m = \frac{\sinh(b )}{ \sqrt{ \sinh^2 b  + e^{-4 j } } } .
    \label{eq:m}
\end{equation}
Average magnetization given by the equation (\ref{eq:m}) has same form as one of the Ising model, but in our model the the dependence of $b$ and $j$ from the temperature and pressure is different from one of the Ising model. 
The fraction of molecules in the HS state is given by the occupation number $n_{HS} = \frac{1 + \left\langle s \right\rangle}{2}$ and the fraction of the molecules in the LS state is $n_{LS} = \frac{1 - \left\langle s \right\rangle}{2}$.
Results for zero pressure, symmetric degeneracies case $g_{+}= g_{-}$ and without phononic part repeat well-known behavior of conventional Ising model.

In Fig.~\ref{fig:Average_magnetization_plots} the dependence of the fraction molecules in the HS state on the temperature for various values of pressure are plotted.
The parameters of the model are chosen to be following: $a_B/a_\epsilon = 0.104$, $a_J/a_\epsilon = 0.0104$, $K_\epsilon/K_B = 1.28$, and $K_\epsilon/K_J = 0.57$ with $K_J a_J^2 = 72 J$.
The thermal behavior of the molecular fraction $n_{HS}(T)$ characterizes the nature of transitions that may be abrupt or gradual, depending on the choice of the values of $T_{eq}$ and $T_{crossover}$.
Under small pressure the cooperativity decreases and the transition becomes less abrupt at higher temperatures.

\section{Average volume and the correlation function}
\label{sec:Volume}

\begin{figure*}[!htb]
    \centering
    \includegraphics[width=0.97\columnwidth]{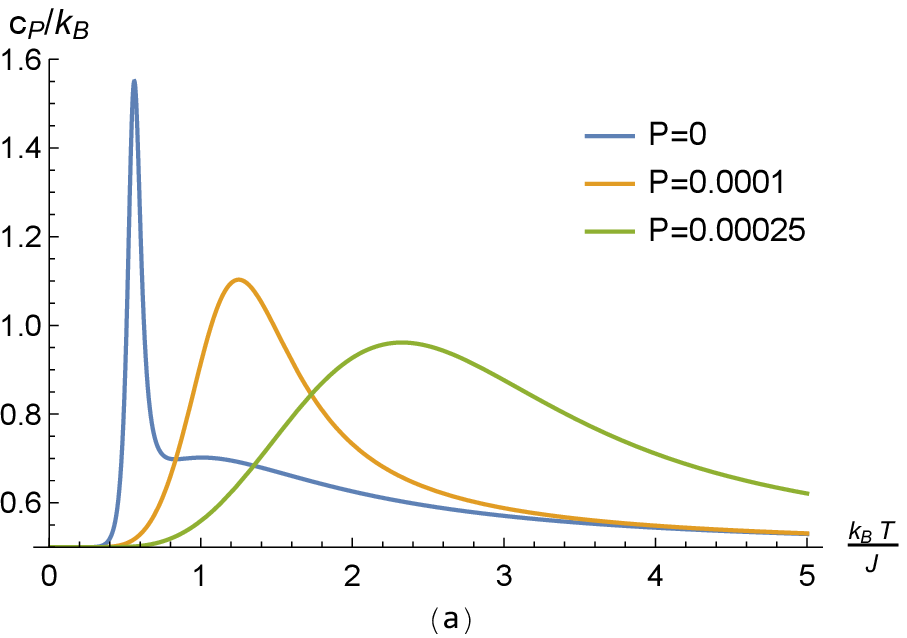}
    \includegraphics[width=0.97\columnwidth]{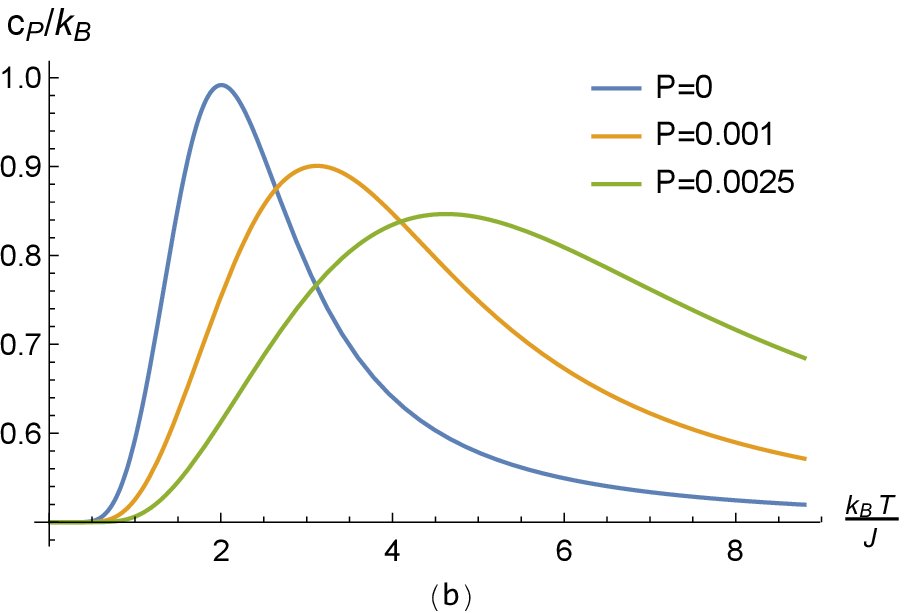}
    \caption{Specific heat capacity per particle $c_P(T)$ as a function of temperature for various pressure (a) $T_{eq} = 0.3 T_{crossover}$. 
    %Parameters are chosen to be same as Fig.~\ref{fig:Average_magnetization_plots}(c).
    %Maximum of the heat capacity is shifted from the maximum of the derivative $\frac{\partial m}{\partial T}$. 
    (b) $T_{eq} = 3 T_{crossover}$. 
    With the pressure increase peak of the specific heat capacity shifts to higher temperatures.
    %Parameters are chosen to be same as in Fig.~\ref{fig:Average_magnetization_plots}(a).
    }
    \label{fig:C_P}
\end{figure*}

Let's calculate average effective volume of the finite molecular chain  
\begin{equation}
    L = \sum_{n=1}^{N-1} \langle x_{n+1} - x_n \rangle = \sum_{n=1}^{N-1} \langle q_n \rangle.
\end{equation}
By the definition, the average distance between the nearest molecules is 
\begin{equation}
    \langle q_n \rangle\! = \! \frac{1}{Z} \sum_{\langle s_1, \ldots, s_N \rangle}\! \iiint dq_1 \cdots dq_{N-1} q_n g_{s_1} \cdots g_{s_N} e^{-\beta E%\left( q_1, \ldots, q_{N-1}, s_1, \ldots, s_N \right)
    } .
\end{equation}
Integrating over the phonon degrees of freedom we get
\begin{equation}
    \langle q_n \rangle = \frac{\sum_{\langle s_1, \ldots, s_N \rangle} \left(  a_{s_ns_{n+1}}  -  \frac{ P}{ K_{s_ns_{n+1}}} \right)  e^{-\beta H_{eff}  }}{\sum_{\langle s_1, \ldots, s_N \rangle}  e^{-\beta H_{eff} }}.
\end{equation}
Suchwise, the effective volume of system (length of molecular chain) is $L = \sum_{n=1}^{N-1} \langle a_{s_ns_{n+1}}  -  \frac{ P}{ K_{s_ns_{n+1}}}  \rangle $.
We rewrite later expression as follows
\begin{multline}
     L = \sum_{n=1}^{N-1} \left( a_\epsilon  -  \frac{ P}{ K_\epsilon} + (a_J -  \frac{ P}{ K_J}  )  \langle s_n s_{n+1}  \rangle + \right. \\ \left. (\frac{a_B}{2} -  \frac{ P}{2 K_B} ) \langle s_n+s_{n+1}  \rangle  \right) .
\end{multline}
Thus, the effective volume of the system is connected with the average magnetization and the correlation function.
The local magnetization may be calculated directly~\cite{gudyma20211d}
\begin{equation}
    \langle \hat{s}_n \rangle = m + \frac{  C_{+-} e ^{-\frac{n-1}{\xi}}  + C_{-+} e ^{-\frac{N-n}{\xi}}  }{{ c_{+} + c_{-} e ^{-\frac{N-1}{\xi}}   } },
    \label{eq:m_exact}
\end{equation}
where coefficients 
\begin{equation}
    C_{-+} = C_{+-} = (m^2-1) (-\sinh w + e^{2j} \sinh b ) ,
\end{equation}
and the correlation length $\xi = - \ln \frac{\lambda_{-}}{\lambda_+}$.
It is easy to see that since $\lambda_-<\lambda_+$, $\xi > 0$.
The average magnetization is
\begin{equation}
    \langle \hat{s} \rangle = m +  \frac{  C_{+-} + C_{-+}}{ N \left( 1 - e ^{-\frac{N}{\xi}} \right) \left(  c_{+} + c_{-} e ^{-\frac{N-1}{\xi}}  \right) } . 
\end{equation}
%In the Sec.~\ref{sec:Hamiltonian} average magnetization already was calculated in the thermodynamic limit.
In the thermodynamic limit, we get classic Ising model magnetization $\langle \hat{s} \rangle = m$.
We note that only average over all spins magnetization coincides with the classic Ising model result, while average of the individual spin is distinct from the classic result due to the system boundary.
We see boundary effects do not vanish even in the thermodynamic limit. 

The local correlation function $G_n(r)$ is~\cite{gudyma20211d}
\begin{multline}
    G_n(r) = \langle \hat{s}_n \hat{s}_{n+r} \rangle = m^2 + (1-m^2) \frac{ c_{+} e ^{-\frac{r}{\xi}} +  c_{-} e ^{-\frac{N-r-1}{\xi}} }{  c_{+} + c_{-} e ^{-\frac{N-1}{\xi}}  }
    \\ +  m C_{+-} \frac{  e ^{-\frac{n-1}{\xi}}-  e ^{-\frac{n-1+r}{\xi}} + e ^{-\frac{N-n-r}{\xi}} -  e ^{-\frac{N-n}{\xi}}  
     }{  c_{+} + c_{-} e ^{-\frac{N-1}{\xi}}  }.
    \label{eq:correlation_function}
\end{multline}
%where coefficients are $C_{-++} = C_{++-}= - C_{--+} = - C_{+--} = m C_{+-}$, $C_{+-+} =  (1-m^2) c_{+}$, and $C_{-+-} = (1-m^2) c_{-}$.
In the thermodynamic limit, we get the correlation function
\begin{equation}
    G(r) = \frac{1}{N} \sum_{n=1}^{N-r-1} \langle \hat{s}_n \hat{s}_{n+r} \rangle = m^2 + (1-m^2) e^{-\frac{r}{\xi}}.
\end{equation}
The average magnetization given by the Eq.~(\ref{eq:m_exact}) and the correlation function given by the Eq.~(\ref{eq:correlation_function}) are exact. 
We see the average correlation function matches with the classic Ising model result~\cite{bellucci2013correlation} in the thermodynamic limit. 
Local correlation function (see Eq.~(\ref{eq:correlation_function})) has information about the edges of the system even in the thermodynamic limit.

Finally, we get the length of chain
\begin{multline}
    L = (N-1) \left( a_\epsilon  -  \frac{ P}{ K_\epsilon} +  (a_B -  \frac{ P}{ K_B} ) m +  (a_J -  \frac{ P}{ K_J}  ) G(1) \right)     
    \\ + (a_B -  \frac{ P}{ K_B} ) \frac{  C_{+-} + C_{-+}  }{{ c_{+} + c_{-} e ^{-\frac{N-1}{\xi}}   } } \left[\frac{1}{1 - e^{-\frac{N}{\xi}}} - \frac{1}{2}(1 + e^{-\frac{N-1}{\xi}}) \right] 
    \\ + (a_J -  \frac{ P}{ K_J}  ) m \frac{C_{+-} ( 1 - e ^{-\frac{1}{\xi}} )}{ c_{+} + c_{-} e ^{-\frac{N-1}{\xi}}  } \left[ \frac{1}{1 - e^{-\frac{N-1}{\xi}}} + \frac{e^{-\frac{N-1}{\xi}}}{1 - e^{\frac{N-1}{\xi}}} \right].
    \label{eq:L_final}
\end{multline}
Expression (\ref{eq:L_final}) is exact and, in the thermodynamic limit, defines average density of the molecular chain $\rho^{-1} = L/N \to a_\epsilon  -  \frac{ P}{ K_\epsilon} +  (a_B -  \frac{ P}{ K_B} ) m +  (a_J -  \frac{ P}{ K_J}  ) G(1)$.

\begin{figure*}[!htb]
    \centering
    \includegraphics[width=0.97\columnwidth]{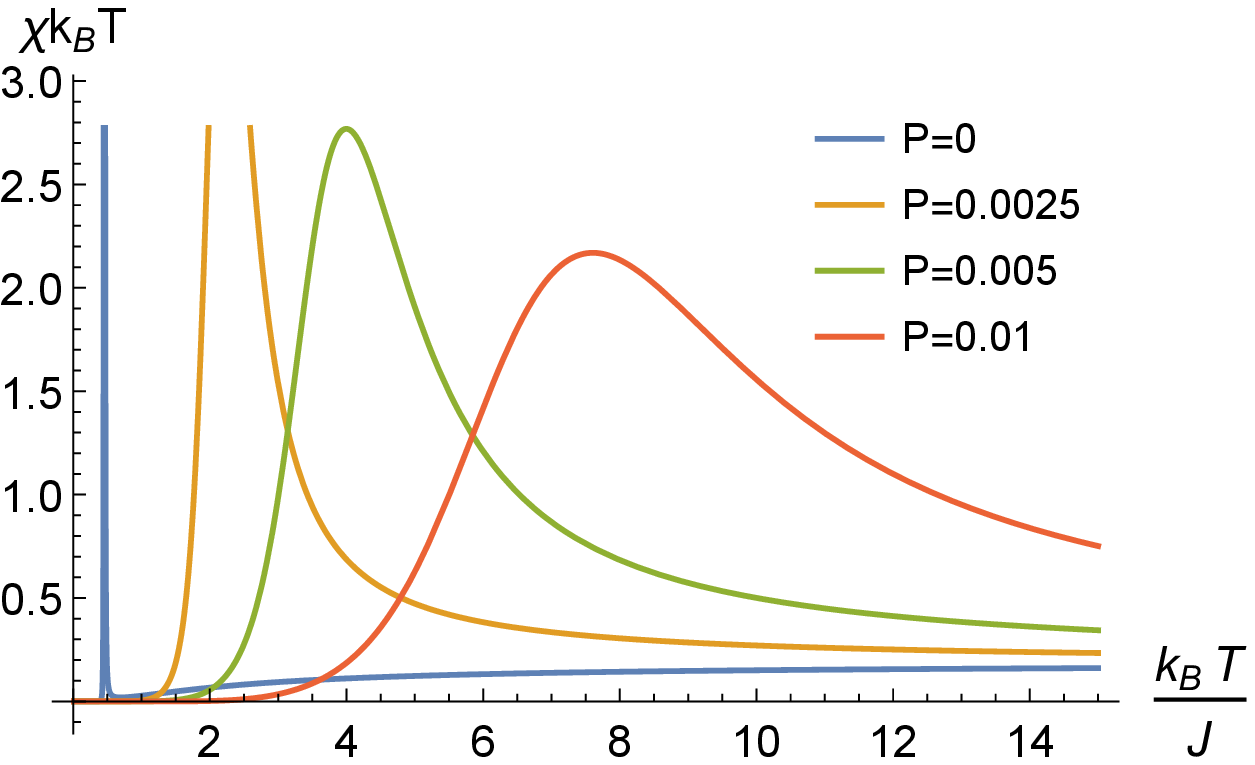}
    \includegraphics[width=0.97\columnwidth]{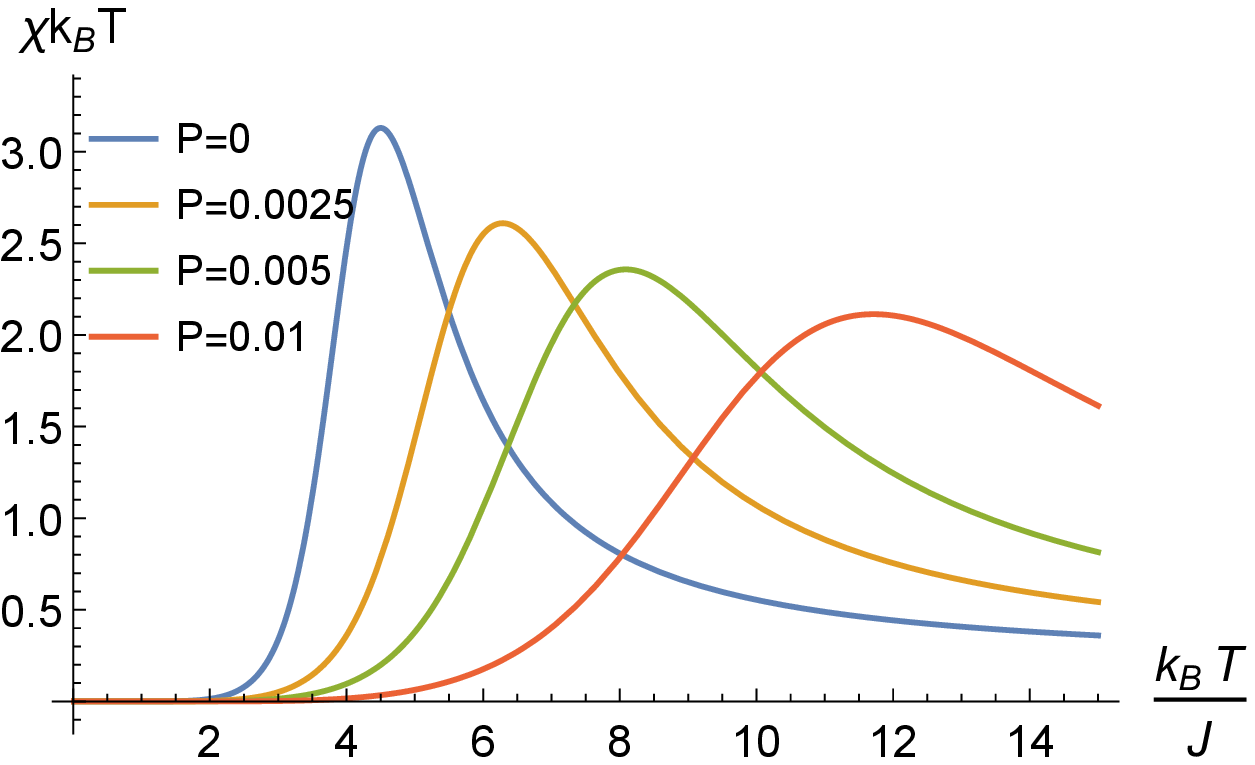}
    \caption{Susceptibility $\chi$ $c_P(T)$ as a function of temperature for various pressure (a) $T_{eq} = 0.3 T_{crossover}$. 
    %Parameters are chosen to be same as Fig.~\ref{fig:Average_magnetization_plots}(c).
    %Maximum of the heat capacity is shifted from the maximum of the derivative $\frac{\partial m}{\partial T}$. 
    (b) $T_{eq} = 3 T_{crossover}$. 
    %Parameters are chosen to be same as in Fig.~\ref{fig:Average_magnetization_plots}(a).
    With the pressure increase peak of the susceptibility shifts to higher temperatures.
    }
    \label{fig:chi}
\end{figure*}

The only approximation we made is the harmonic approximation of the interparticle potential.
This approximation should be valid when the displacement of particles from the equilibrium distances is small.
Applied external pressure clearly reduces the distances between the particles and at some point harmonic approximation loose its validity.
Eq.~(\ref{eq:L_final}) gives us some understanding of the harmonic approximation limits.
The volume of the system should be positive, therefore we external pressure should satisfy following condition $P \ll P_{\epsilon} = a_\epsilon K_\epsilon$.

\section{Specific heat capacity and susceptibility}
\label{sec:Heat_capacity}

The specific heat capacity is one of the most important thermodynamic characteristic of the system which can be easily measured experimentally.
We consider a system under a constant pressure, and therefore the volume of the system changes. 
The internal energy per particle is $ \langle E \rangle = - \frac{\partial}{\partial \beta} \ln Z$
\begin{multline}
    \langle E \rangle = E_0 +  \frac{1}{2} N k_B T  - \sum_{n=1}^{N-1} (J + a_J P - \frac{  P^2}{2 K_J }) \left\langle s_n s_{n+1} \right\rangle 
    \\- \sum_{n=1}^{N}  (B + a_B P - \frac{ P^2}{2 K_B }) \langle s_n \rangle . 
\end{multline}
%Therefore the enthalpy is $H = \langle E \rangle + P V$, where the volume $V$ is given by the Eq.~(\ref{eq:L_final}). 
And the heat capacity per particle $c_P = \frac{1}{N} \frac{\partial \langle E \rangle}{\partial T}$ in the thermodynamic limit can be written in the following way
\begin{multline}
    c_P = \frac{1}{2} k_B - \left(B + a_B P - \frac{ P^2}{2 K_B } \right) \frac{\partial m}{\partial T} 
    \\ - \left(J + a_J P - \frac{  P^2}{2 K_J } \right)  \frac{\partial }{\partial T} G(1) .
\end{multline}
Specific heat capacities per particle for small pressures are given in Fig.~\ref{fig:C_P}.
Parameters of the model in Figs.~\ref{fig:C_P} and \ref{fig:chi} are the same as in Fig.~\ref{fig:Average_magnetization_plots}.
One-dimensional systems demonstrate two-peak specific heat capacity thermal behavior on experiments~\cite{Garcia2015Two-Step}.
Our model captures this phenomenon at small pressure in the abrupt crossover regime.
Main peak is associated with the Schottky anomaly. 
This result may be expected as we was demonstrated the exact mapping onto the Ising-like system with the Hamiltonian (\ref{eq:H_eff}).
Such behavior appear as a result of the initial assumption about the nature of iron(II) materials that only two lowest single-molecule levels (denoted LS and HS) are relevant for the description of the system.  
With the increase of pressure the main peak of specific heat capacity become more broad and shifts to higher temperatures.
Such behaviour quickly disappears with pressure increase.

The susceptibility $\chi = 2 \frac{\partial n_{HS}}{\partial B}$ is
\begin{equation}
    \chi = \frac{1}{k_B T} \frac{\cosh(b ) e^{-4 j } }{ \left( \sinh^2 b  + e^{-4 j } \right)^{\frac{3}{2}} }. 
\end{equation}
The susceptibility as a function of $k_{B} T/J$ under various pressure is shown in Fig.~\ref{fig:chi}.
Similarly to the specific heat capacity, with the pressure increase the peak of the susceptibility shifts to higher temperatures and becomes more broad.

\section{Spin crossover under the pressure}
\label{sec:Pressure}

%One-dimensional systems do not experience true phase transition at any finite temperature, but nevertheless it is useful to introduce the notation of critical temperature $T_c$.
%We adopt this notation from the similar models in higher dimensions where phase transition occur. 
%We define critical temperature from the model analysis in the mean-field approximation.
%The Curie-Weiss equation has form $m = \tanh(z j m + b)$, where $z$ is the number of closest neighbors. 
%For the one dimensional Ising model $z = 2$.
%At the $b=0$ (or equivalently $B = - (\frac{1}{2} \ln g + \delta b) k_B T $) the Curie-Weiss equation gives condition for the critical temperature: $z j \geq 1$. 
%Therefore 
%\begin{equation}
%    T_c = \frac{z ( J +  P a_J - \frac{  P^2}{2 K_J } ) }{k_B (1 - \delta j)}.
%\end{equation}
%First, we notice the phonon interactions change the critical temperature for high-dimensional systems.
%This effect may be important if the elastic constants of the pseudo-states interactions differs a lot. 
%Moreover, if $\delta j > 1$, $T_c$ is negative meaning there is no critical temperature in system. 

In our previous paper~\cite{gudyma20211d} we investigated regimes of gradual and abrupt crossover under zero pressure $P=0$.
We introduced two characteristic values of the system, namely the equilibrium $T_{eq}$ and the crossover temperature $T_{crossover}$.
We demonstrated that if $T_{eq} < T_{crossover}$ the crossover is abrupt and some thermal quantities resemble ones for the phase transition, and if $T_{eq}<0$ or $T_{eq} > T_{crossover}$ the crossover is gradual.
Here our goal is to explore what changes crossover undergo in the case $P \neq 0$.
At zero temperature system always stays in ordered phase which is defined by the sign of the effective field 
\begin{equation}
    n_{HS}(T \to 0) = \frac{1}{2}(1 + \sign(B + a_B P - \frac{ P^2}{2 K_B })).    
\end{equation}
Usually, in Fe (II) compounds $B>0$ and LS is lower than HS state under zero pressure, thus $n_{HS}(T \to 0) = 0$.
Nevertheless for large pressure $B + a_B P - \frac{ P^2}{2 K_B } < 0$ and therefore crossover starts from $n_{HS}(T \to 0) = 1$.
At the same time at high temperatures occupation numbers are
\begin{equation}
    n_{HS}(T \to \infty) =\frac{1}{2} + \frac{1}{2} \frac{\sinh\left(  \frac{1}{2} \ln g + \delta b \right)}{ \sqrt{ \sinh^2 \left(  \frac{1}{2} \ln g + \delta b   \right) + e^{ - 4 \delta j} } } .
    \label{eq:m_high_T_limit}
\end{equation}

\begin{figure*}[!htb]
    \centering
    \includegraphics[width=0.66\columnwidth]{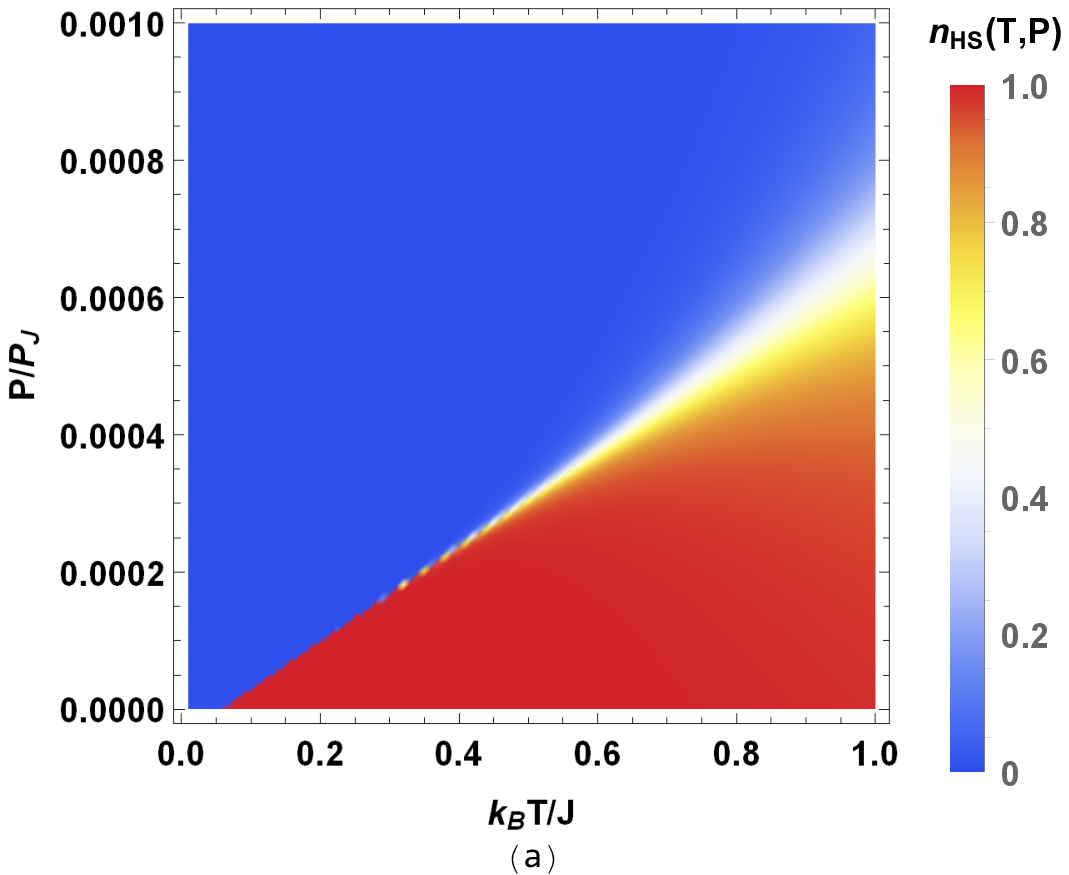}
    \includegraphics[width=0.66\columnwidth]{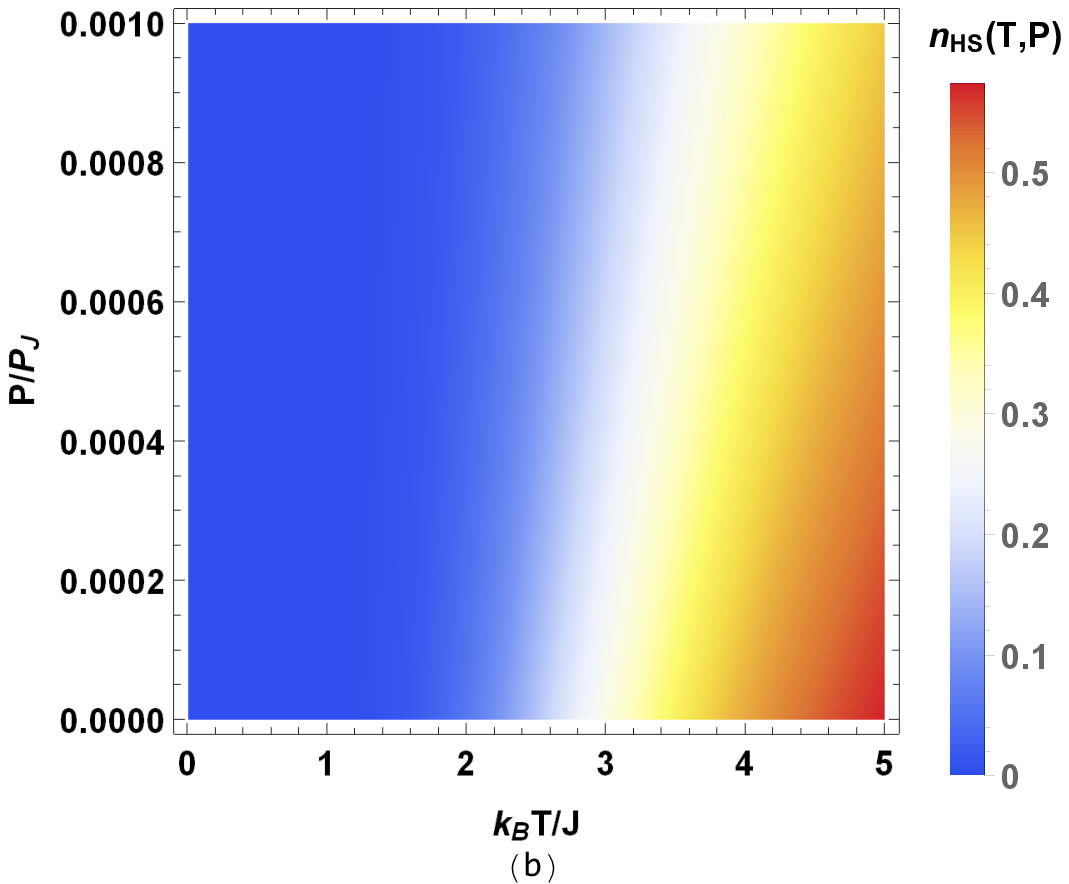}
    \includegraphics[width=0.66\columnwidth]{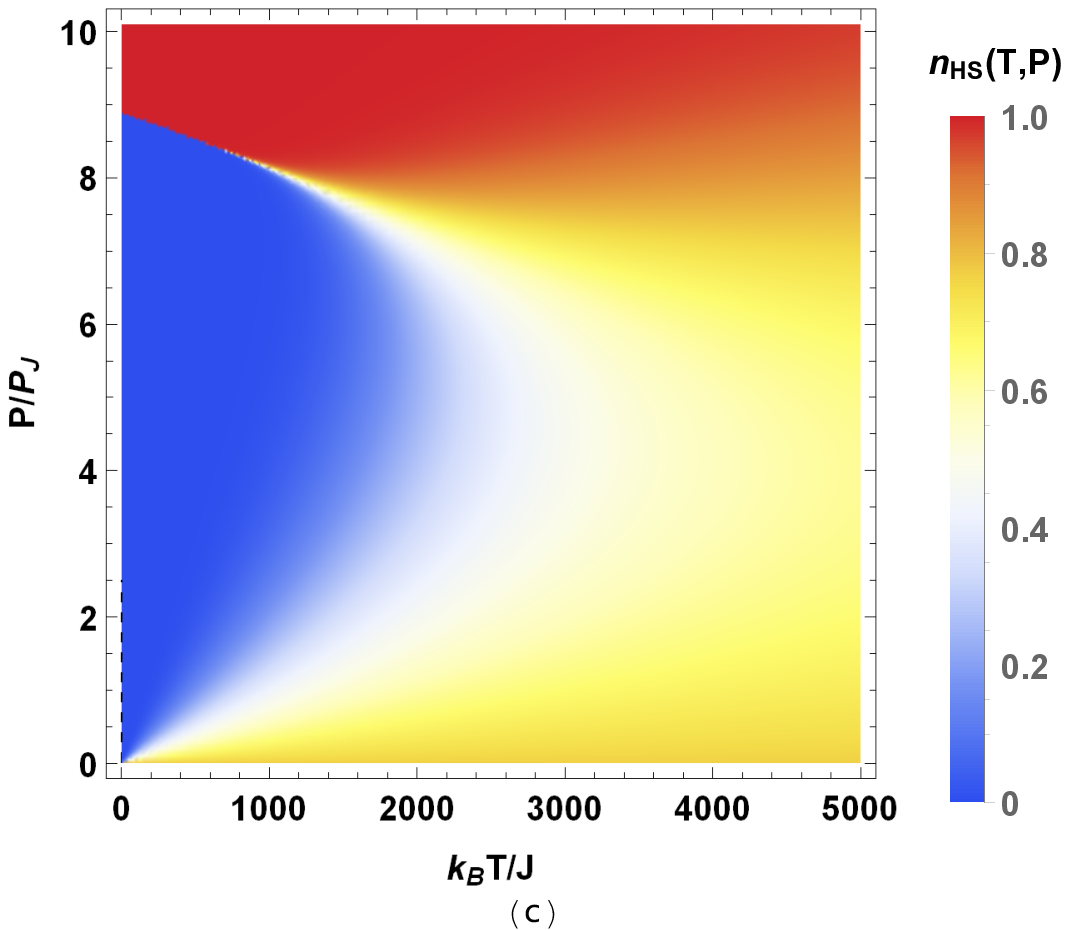}
    \caption{Phase diagram of the average occupation number $n_{HS}(T, P)$. 
    (a) small $(T,P)$ region for $T_{eq}(P=0) < T_{crosover}$,
    (b) small $(T,P)$ region for $T_{eq}(P=0) > T_{crosover}$,
    (c) large scale $(T,P)$ dependence.
    %Dashed line indicates condition $T=T_{eq}$. 
    Colors in the vertical column on the right represent the fraction of high-spin molecules.
    }
    \label{fig:Average_magnetization_phase_diagram}
\end{figure*}

We introduce the equilibrium temperature $T_{eq}$ as a temperature when pseudo-spin states have equal occupations $n_{HS} = 1/2$.
Often the equilibrium temperature is denoted as $T_{1/2}$.
This happens when the effective field vanishes, i.e. $b = 0$.
In doing so, one gets the following expression for the equilibrium temperature $T_{eq}$ as a function of the pressure 
\begin{equation}
    T_{eq}(P) = - \frac{ B -  P a_b + \frac{  P^2}{2 K_B }}{k_B \left( \frac{1}{2} \ln g + \delta b \right) } .
\end{equation}
We note that for certain values of the external field $B$, pressure $P$ and pseudo-spin degeneracies $g$ the equilibrium temperature $T_{eq}$ can be negative what means that for given field and degeneracies there is no such temperature that pseudo-spin states would have equal occupations.

Let's find a temperature $T_0$ for which the occupation number is maximal. 
This temperature should be a solution of the equation
\begin{equation}
    \left. \frac{\partial n_{HS}}{\partial T} \right|_{T=T_0} = 0. 
    \label{eq:phase_transition_condition}
\end{equation}
Therefore
\begin{equation}
    T_0 =  \frac{T_{eq}(P)}{1 - \frac{1}{\frac{1}{2} \ln g + \delta b } \arctanh \left( \frac{ k_B T_{eq}(P) }{ 2 \left( J +  P a_J - \frac{  P^2}{2 K_J } \right)} \left(  \frac{1}{2} \ln g + \delta b   \right) \right)}  .
    \label{eq:T0}
\end{equation}
We call the maximal equilibrium temperature $T_{eq}$ at which Eq.~(\ref{eq:phase_transition_condition}) has finite solutions for the $T_0$ as the crossover temperature.
The derivative $\partial m / \partial T$ is always positive and the occupation number $n_{HS}$ is a monotonous function of temperature when $T_{eq}< T_{crossover}$.
Thus we get the crossover temperature 
\begin{equation}
    T_{crossover} (P) = \frac{2 \left( J +  P a_J - \frac{  P^2}{2 K_J } \right)}{k_B} \frac{\tanh \left( \frac{1}{2} \ln g + \delta b \right) }{\frac{1}{2} \ln g + \delta b}.
    \label{eq:Tcrossover}
\end{equation}
The crossover temperature $T_{crossover} (P) > T_{crossover} (0)$ when the pressure $P < P_J =  2 K_J a_J$, and $T_{crossover} (P) < T_{crossover} (0)$ when $P > P_J$.
Therefore we shall observe abrupt crossover when $T_{crossover}(P)>0$ and $T_{eq}(P) < T_{crossover} (P)$, and gradual crossover otherwise.

The resulting phase diagram for the spin crossover is presented in Fig.~\ref{fig:Average_magnetization_phase_diagram}.
In left and central panels (a)-(b) occupation number is depicted for temperature and pressure values close zero.
A spin crossover phase diagram, where the HS fraction is indicated by color, is shown in a wide range of temperature and pressure variations in Fig.~\ref{fig:Average_magnetization_phase_diagram}c. 
The diagram shows regions of the HS paramagnetic phases under high pressure, and the LS diamagnetic phase at relatively low temperature and pressure.
We observe two regions with abrupt HS--LS transitions: the region near $P=0$ and $P=P_J$.
For $T_{eq}(0) > T_{crosover}(0)$, it can be seen that no sharp discontinuous changes in $n_{HS}$, therefore in structural or optical properties, should be expected to occur across this spin crossover.
As the pressure increases, the width of the SCO region is broadened, the sharp spin transition becomes a smoother and broader SCO.
Evidently, system undergoes a sharp HS-LS transition with a very narrow SCO region at low temperature when $T_{eq} < T_{crosover}$.

\section{Summary and conclusions}
\label{sec:Summary}

The aim of this paper was to give a thorough discussion of thermodynamic properties of the one-dimensional spin-crossover systems being a subject of a constant pressure.
We start with the exact microscopic Hamiltonian which consists of sum of the pair intermolecular potentials. 
In the harmonic approximation we demonstrate exact mapping to the Ising-like Hamiltonian with temperature dependent effective parameters of the model, namely the reference energy, ferromagnetic constant and magnetic field.
For this purpose, the transfer-matrix method was transformed to form that addresses free-boundary case.
The elaborated rigorous procedure has enabled us to derive exact results for the basic thermodynamic quantities and pair correlation function.
In framework this approach we show that the degeneracy of the levels, elastic interaction and pressure renormalize the parameters of the effective Ising model. 
We analyze regimes of the HS--LS crossover and identify regions of parameters where the crossover abrupt or gradual and show how pressure effects on the location and size of the transition.
In the next works we are planning to extend our results to higher dimensions and experimental situations.

\section*{Data Availability}

The data that support the findings of this study are available from the corresponding author upon reasonable request.

% If in two-column mode, this environment will change to single-column format so that long equations can be displayed. 
%\begin{widetext}
%$$\mbox{put long equation here}$$
%\end{widetext}

% Figures should be put into the text as floats. 
% Use the graphics or graphicx packages (distributed with LaTeX2e).
% Put the label that you will use with \ref{} command in the braces of the \label{} command.
%
% \begin{figure}
% \end{figure}

% \begin{table}
% \end{tabular}
% \end{table}

%\begin{acknowledgments}
% Put your acknowledgments here.
%\end{acknowledgments}


\section*{References}
\bibliography{main}

\providecommand{\noopsort}[1]{}\providecommand{\singleletter}[1]{#1}%
\begin{thebibliography}{43}%
\makeatletter
\providecommand \@ifxundefined [1]{%
 \@ifx{#1\undefined}
}%
\providecommand \@ifnum [1]{%
 \ifnum #1\expandafter \@firstoftwo
 \else \expandafter \@secondoftwo
 \fi
}%
\providecommand \@ifx [1]{%
 \ifx #1\expandafter \@firstoftwo
 \else \expandafter \@secondoftwo
 \fi
}%
\providecommand \natexlab [1]{#1}%
\providecommand \enquote  [1]{``#1''}%
\providecommand \bibnamefont  [1]{#1}%
\providecommand \bibfnamefont [1]{#1}%
\providecommand \citenamefont [1]{#1}%
\providecommand \href@noop [0]{\@secondoftwo}%
\providecommand \href [0]{\begingroup \@sanitize@url \@href}%
\providecommand \@href[1]{\@@startlink{#1}\@@href}%
\providecommand \@@href[1]{\endgroup#1\@@endlink}%
\providecommand \@sanitize@url [0]{\catcode `\\12\catcode `\$12\catcode
  `\&12\catcode `\#12\catcode `\^12\catcode `\_12\catcode `\%12\relax}%
\providecommand \@@startlink[1]{}%
\providecommand \@@endlink[0]{}%
\providecommand \url  [0]{\begingroup\@sanitize@url \@url }%
\providecommand \@url [1]{\endgroup\@href {#1}{\urlprefix }}%
\providecommand \urlprefix  [0]{URL }%
\providecommand \Eprint [0]{\href }%
\providecommand \doibase [0]{http://dx.doi.org/}%
\providecommand \selectlanguage [0]{\@gobble}%
\providecommand \bibinfo  [0]{\@secondoftwo}%
\providecommand \bibfield  [0]{\@secondoftwo}%
\providecommand \translation [1]{[#1]}%
\providecommand \BibitemOpen [0]{}%
\providecommand \bibitemStop [0]{}%
\providecommand \bibitemNoStop [0]{.\EOS\space}%
\providecommand \EOS [0]{\spacefactor3000\relax}%
\providecommand \BibitemShut  [1]{\csname bibitem#1\endcsname}%
\let\auto@bib@innerbib\@empty
%</preamble>
\bibitem [{\citenamefont {Jureschi}\ \emph {et~al.}(2015)\citenamefont
  {Jureschi}, \citenamefont {Linares}, \citenamefont {Rotaru}, \citenamefont
  {Ritti}, \citenamefont {Parlier}, \citenamefont {D{\^\i}rtu}, \citenamefont
  {Wolff},\ and\ \citenamefont {Garcia}}]{jureschi2015pressure}%
  \BibitemOpen
  \bibfield  {author} {\bibinfo {author} {\bibfnamefont {C.~M.}\ \bibnamefont
  {Jureschi}}, \bibinfo {author} {\bibfnamefont {J.}~\bibnamefont {Linares}},
  \bibinfo {author} {\bibfnamefont {A.}~\bibnamefont {Rotaru}}, \bibinfo
  {author} {\bibfnamefont {M.~H.}\ \bibnamefont {Ritti}}, \bibinfo {author}
  {\bibfnamefont {M.}~\bibnamefont {Parlier}}, \bibinfo {author} {\bibfnamefont
  {M.~M.}\ \bibnamefont {D{\^\i}rtu}}, \bibinfo {author} {\bibfnamefont
  {M.}~\bibnamefont {Wolff}}, \ and\ \bibinfo {author} {\bibfnamefont
  {Y.}~\bibnamefont {Garcia}},\ }\bibfield  {title} {\enquote {\bibinfo {title}
  {Pressure sensor via optical detection based on a 1d spin transition
  coordination polymer},}\ }\href {\doibase http://doi.org/10.3390/s150202388}
  {\bibfield  {journal} {\bibinfo  {journal} {Sensors}\ }\textbf {\bibinfo
  {volume} {15}},\ \bibinfo {pages} {2388--2398} (\bibinfo {year}
  {2015})}\BibitemShut {NoStop}%
\bibitem [{\citenamefont {Gudyma}, \citenamefont {Enachescu},\ and\
  \citenamefont {Maksymov}(2015)}]{gudyma2015kinetics}%
  \BibitemOpen
  \bibfield  {author} {\bibinfo {author} {\bibfnamefont {{\relax
  Iu}.}~\bibnamefont {Gudyma}}, \bibinfo {author} {\bibfnamefont
  {C.}~\bibnamefont {Enachescu}}, \ and\ \bibinfo {author} {\bibfnamefont
  {A.}~\bibnamefont {Maksymov}},\ }\bibfield  {title} {\enquote {\bibinfo
  {title} {Kinetics of nonequilibrium transition in spin-crossover
  compounds},}\ }in\ \href {\doibase
  https://doi.org/10.1007/978-3-319-06611-0_29} {\emph {\bibinfo {booktitle}
  {Nanocomposites, Nanophotonics, Nanobiotechnology, and Applications}}}\
  (\bibinfo  {publisher} {Springer},\ \bibinfo {year} {2015})\ pp.\ \bibinfo
  {pages} {375--401}\BibitemShut {NoStop}%
\bibitem [{\citenamefont {Halcrow}(2013)}]{halcrow2013spin}%
  \BibitemOpen
  \bibinfo {editor} {\bibfnamefont {M.~A.}\ \bibnamefont {Halcrow}},\ ed.,\
  \href@noop {} {\emph {\bibinfo {title} {Spin-crossover materials: properties
  and applications}}}\ (\bibinfo  {publisher} {John Wiley \& Sons},\ \bibinfo
  {year} {2013})\BibitemShut {NoStop}%
\bibitem [{\citenamefont {G{\"u}tlich}, \citenamefont {Hauser},\ and\
  \citenamefont {Spiering}(1994)}]{3d43d7-1}%
  \BibitemOpen
  \bibfield  {author} {\bibinfo {author} {\bibfnamefont {P.}~\bibnamefont
  {G{\"u}tlich}}, \bibinfo {author} {\bibfnamefont {A.}~\bibnamefont {Hauser}},
  \ and\ \bibinfo {author} {\bibfnamefont {H.}~\bibnamefont {Spiering}},\
  }\bibfield  {title} {\enquote {\bibinfo {title} {Thermal and optical
  switching of iron (ii) complexes},}\ }\href {\doibase
  https://doi.org/10.1002/anie.199420241} {\bibfield  {journal} {\bibinfo
  {journal} {Angewandte Chemie International Edition in English}\ }\textbf
  {\bibinfo {volume} {33}},\ \bibinfo {pages} {2024--2054} (\bibinfo {year}
  {1994})}\BibitemShut {NoStop}%
\bibitem [{\citenamefont {Coronado}(2020)}]{coronado2020molecular}%
  \BibitemOpen
  \bibfield  {author} {\bibinfo {author} {\bibfnamefont {E.}~\bibnamefont
  {Coronado}},\ }\bibfield  {title} {\enquote {\bibinfo {title} {Molecular
  magnetism: from chemical design to spin control in molecules, materials and
  devices},}\ }\href {\doibase 10.1038/s41578-019-0146-8} {\bibfield  {journal}
  {\bibinfo  {journal} {Nature Reviews Materials}\ }\textbf {\bibinfo {volume}
  {5}},\ \bibinfo {pages} {87--104} (\bibinfo {year} {2020})}\BibitemShut
  {NoStop}%
\bibitem [{\citenamefont {G{\"u}tlich}, \citenamefont {Goodwin},\ and\
  \citenamefont {Garcia}(2004)}]{3d43d7-3}%
  \BibitemOpen
  \bibfield  {author} {\bibinfo {author} {\bibfnamefont {P.}~\bibnamefont
  {G{\"u}tlich}}, \bibinfo {author} {\bibfnamefont {H.~A.}\ \bibnamefont
  {Goodwin}}, \ and\ \bibinfo {author} {\bibfnamefont {Y.}~\bibnamefont
  {Garcia}},\ }\href@noop {} {\emph {\bibinfo {title} {Spin crossover in
  transition metal compounds I}}},\ Vol.~\bibinfo {volume} {1}\ (\bibinfo
  {publisher} {Springer Science \& Business Media},\ \bibinfo {year}
  {2004})\BibitemShut {NoStop}%
\bibitem [{\citenamefont {Bousseksou}\ \emph {et~al.}(2011)\citenamefont
  {Bousseksou}, \citenamefont {Moln{\'a}r}, \citenamefont {Salmon},\ and\
  \citenamefont {Nicolazzi}}]{3d43d7-5}%
  \BibitemOpen
  \bibfield  {author} {\bibinfo {author} {\bibfnamefont {A.}~\bibnamefont
  {Bousseksou}}, \bibinfo {author} {\bibfnamefont {G.}~\bibnamefont
  {Moln{\'a}r}}, \bibinfo {author} {\bibfnamefont {L.}~\bibnamefont {Salmon}},
  \ and\ \bibinfo {author} {\bibfnamefont {W.}~\bibnamefont {Nicolazzi}},\
  }\bibfield  {title} {\enquote {\bibinfo {title} {Molecular spin crossover
  phenomenon: recent achievements and prospects},}\ }\href {\doibase
  10.1039/C1CS15042A} {\bibfield  {journal} {\bibinfo  {journal} {Chemical
  Society Reviews}\ }\textbf {\bibinfo {volume} {40}},\ \bibinfo {pages}
  {3313--3335} (\bibinfo {year} {2011})}\BibitemShut {NoStop}%
\bibitem [{\citenamefont {Garcia}\ \emph {et~al.}(2000)\citenamefont {Garcia},
  \citenamefont {Ksenofontov}, \citenamefont {Levchenko},\ and\ \citenamefont
  {Gütlich}}]{Experiment1d-1}%
  \BibitemOpen
  \bibfield  {author} {\bibinfo {author} {\bibfnamefont {Y.}~\bibnamefont
  {Garcia}}, \bibinfo {author} {\bibfnamefont {V.}~\bibnamefont {Ksenofontov}},
  \bibinfo {author} {\bibfnamefont {G.}~\bibnamefont {Levchenko}}, \ and\
  \bibinfo {author} {\bibfnamefont {P.}~\bibnamefont {Gütlich}},\ }\bibfield
  {title} {\enquote {\bibinfo {title} {Pressure effect on a novel spin
  transition polymeric chain compound},}\ }\href {\doibase 10.1039/B003794J}
  {\bibfield  {journal} {\bibinfo  {journal} {J. Mater. Chem.}\ }\textbf
  {\bibinfo {volume} {10}},\ \bibinfo {pages} {2274--2276} (\bibinfo {year}
  {2000})}\BibitemShut {NoStop}%
\bibitem [{\citenamefont {Sugahara}\ \emph {et~al.}(2017)\citenamefont
  {Sugahara}, \citenamefont {Kamebuchi}, \citenamefont {Okazawa}, \citenamefont
  {Enomoto},\ and\ \citenamefont {Kojima}}]{sugahara2017control}%
  \BibitemOpen
  \bibfield  {author} {\bibinfo {author} {\bibfnamefont {A.}~\bibnamefont
  {Sugahara}}, \bibinfo {author} {\bibfnamefont {H.}~\bibnamefont {Kamebuchi}},
  \bibinfo {author} {\bibfnamefont {A.}~\bibnamefont {Okazawa}}, \bibinfo
  {author} {\bibfnamefont {M.}~\bibnamefont {Enomoto}}, \ and\ \bibinfo
  {author} {\bibfnamefont {N.}~\bibnamefont {Kojima}},\ }\bibfield  {title}
  {\enquote {\bibinfo {title} {Control of spin-crossover phenomena in
  one-dimensional triazole-coordinated iron (ii) complexes by means of
  functional counter ions},}\ }\href {\doibase 10.3390/inorganics5030050}
  {\bibfield  {journal} {\bibinfo  {journal} {Inorganics}\ }\textbf {\bibinfo
  {volume} {5}},\ \bibinfo {pages} {50} (\bibinfo {year} {2017})}\BibitemShut
  {NoStop}%
\bibitem [{\citenamefont {Nebbali}\ \emph {et~al.}(2018)\citenamefont
  {Nebbali}, \citenamefont {Mekuimemba}, \citenamefont {Charles}, \citenamefont
  {Yefsah}, \citenamefont {Chastanet}, \citenamefont {Mota}, \citenamefont
  {Colacio},\ and\ \citenamefont {Triki}}]{nebbali2018one}%
  \BibitemOpen
  \bibfield  {author} {\bibinfo {author} {\bibfnamefont {K.}~\bibnamefont
  {Nebbali}}, \bibinfo {author} {\bibfnamefont {C.~D.}\ \bibnamefont
  {Mekuimemba}}, \bibinfo {author} {\bibfnamefont {C.}~\bibnamefont {Charles}},
  \bibinfo {author} {\bibfnamefont {S.}~\bibnamefont {Yefsah}}, \bibinfo
  {author} {\bibfnamefont {G.}~\bibnamefont {Chastanet}}, \bibinfo {author}
  {\bibfnamefont {A.~J.}\ \bibnamefont {Mota}}, \bibinfo {author}
  {\bibfnamefont {E.}~\bibnamefont {Colacio}}, \ and\ \bibinfo {author}
  {\bibfnamefont {S.}~\bibnamefont {Triki}},\ }\bibfield  {title} {\enquote
  {\bibinfo {title} {One-dimensional thiocyanato-bridged fe (ii) spin crossover
  cooperative polymer with unusual fen5s coordination sphere},}\ }\href
  {\doibase 10.1021/acs.inorgchem.8b02061} {\bibfield  {journal} {\bibinfo
  {journal} {Inorganic chemistry}\ }\textbf {\bibinfo {volume} {57}},\ \bibinfo
  {pages} {12338--12346} (\bibinfo {year} {2018})}\BibitemShut {NoStop}%
\bibitem [{\citenamefont {Wolny}\ \emph {et~al.}(2020)\citenamefont {Wolny},
  \citenamefont {Hochd{\"o}rffer}, \citenamefont {Sadashivaiah}, \citenamefont
  {Auerbach}, \citenamefont {Jenni}, \citenamefont {Scherthan}, \citenamefont
  {Li}, \citenamefont {von Malotki}, \citenamefont {Wille}, \citenamefont
  {Rentschler} \emph {et~al.}}]{wolny2020vibrational}%
  \BibitemOpen
  \bibfield  {author} {\bibinfo {author} {\bibfnamefont {J.~A.}\ \bibnamefont
  {Wolny}}, \bibinfo {author} {\bibfnamefont {T.}~\bibnamefont
  {Hochd{\"o}rffer}}, \bibinfo {author} {\bibfnamefont {S.}~\bibnamefont
  {Sadashivaiah}}, \bibinfo {author} {\bibfnamefont {H.}~\bibnamefont
  {Auerbach}}, \bibinfo {author} {\bibfnamefont {K.}~\bibnamefont {Jenni}},
  \bibinfo {author} {\bibfnamefont {L.}~\bibnamefont {Scherthan}}, \bibinfo
  {author} {\bibfnamefont {A.-M.}\ \bibnamefont {Li}}, \bibinfo {author}
  {\bibfnamefont {C.}~\bibnamefont {von Malotki}}, \bibinfo {author}
  {\bibfnamefont {H.-C.}\ \bibnamefont {Wille}}, \bibinfo {author}
  {\bibfnamefont {E.}~\bibnamefont {Rentschler}},  \emph {et~al.},\ }\bibfield
  {title} {\enquote {\bibinfo {title} {Vibrational properties of 1d-and 3d
  polynuclear spin crossover fe (ii) urea-triazoles polymer chains and
  quantification of intrachain cooperativity},}\ }\href {\doibase
  10.1088/1361-648X/aba71d} {\bibfield  {journal} {\bibinfo  {journal} {Journal
  of Physics: Condensed Matter}\ }\textbf {\bibinfo {volume} {33}},\ \bibinfo
  {pages} {034004} (\bibinfo {year} {2020})}\BibitemShut {NoStop}%
\bibitem [{\citenamefont {Weselski}\ \emph {et~al.}(2017)\citenamefont
  {Weselski}, \citenamefont {Książek}, \citenamefont {Kusz}, \citenamefont
  {Białońska}, \citenamefont {Paliwoda}, \citenamefont {Hanfland},
  \citenamefont {Rudolf}, \citenamefont {Ciunik},\ and\ \citenamefont
  {Bronisz}}]{Experiment1d-Bronisz}%
  \BibitemOpen
  \bibfield  {author} {\bibinfo {author} {\bibfnamefont {M.}~\bibnamefont
  {Weselski}}, \bibinfo {author} {\bibfnamefont {M.}~\bibnamefont {Książek}},
  \bibinfo {author} {\bibfnamefont {J.}~\bibnamefont {Kusz}}, \bibinfo {author}
  {\bibfnamefont {A.}~\bibnamefont {Białońska}}, \bibinfo {author}
  {\bibfnamefont {D.}~\bibnamefont {Paliwoda}}, \bibinfo {author}
  {\bibfnamefont {M.}~\bibnamefont {Hanfland}}, \bibinfo {author}
  {\bibfnamefont {M.~F.}\ \bibnamefont {Rudolf}}, \bibinfo {author}
  {\bibfnamefont {Z.}~\bibnamefont {Ciunik}}, \ and\ \bibinfo {author}
  {\bibfnamefont {R.}~\bibnamefont {Bronisz}},\ }\bibfield  {title} {\enquote
  {\bibinfo {title} {Evidence of ligand elasticity occurring in temperature-,
  light-, and pressure-induced spin crossover in 1d coordination polymers
  [fe(3ditz)3]x2 (x = clo4–, bf4–)},}\ }\href {\doibase
  https://doi.org/10.1002/ejic.201601399} {\bibfield  {journal} {\bibinfo
  {journal} {European Journal of Inorganic Chemistry}\ }\textbf {\bibinfo
  {volume} {2017}},\ \bibinfo {pages} {1171--1179} (\bibinfo {year}
  {2017})}\BibitemShut {NoStop}%
\bibitem [{\citenamefont {Dîrtu}\ \emph {et~al.}(2015)\citenamefont {Dîrtu},
  \citenamefont {Schmit}, \citenamefont {Naik}, \citenamefont {Rusu},
  \citenamefont {Rotaru}, \citenamefont {Rackwitz}, \citenamefont {Wolny},
  \citenamefont {Schünemann}, \citenamefont {Spinu},\ and\ \citenamefont
  {Garcia}}]{Garcia2015Two-Step}%
  \BibitemOpen
  \bibfield  {author} {\bibinfo {author} {\bibfnamefont {M.~M.}\ \bibnamefont
  {Dîrtu}}, \bibinfo {author} {\bibfnamefont {F.}~\bibnamefont {Schmit}},
  \bibinfo {author} {\bibfnamefont {A.~D.}\ \bibnamefont {Naik}}, \bibinfo
  {author} {\bibfnamefont {I.}~\bibnamefont {Rusu}}, \bibinfo {author}
  {\bibfnamefont {A.}~\bibnamefont {Rotaru}}, \bibinfo {author} {\bibfnamefont
  {S.}~\bibnamefont {Rackwitz}}, \bibinfo {author} {\bibfnamefont {J.~A.}\
  \bibnamefont {Wolny}}, \bibinfo {author} {\bibfnamefont {V.}~\bibnamefont
  {Schünemann}}, \bibinfo {author} {\bibfnamefont {L.}~\bibnamefont {Spinu}},
  \ and\ \bibinfo {author} {\bibfnamefont {Y.}~\bibnamefont {Garcia}},\
  }\bibfield  {title} {\enquote {\bibinfo {title} {Two-step spin transition in
  a 1d feii 1,2,4-triazole chain compound},}\ }\href {\doibase
  https://doi.org/10.1002/chem.201406231} {\bibfield  {journal} {\bibinfo
  {journal} {Chemistry – A European Journal}\ }\textbf {\bibinfo {volume}
  {21}},\ \bibinfo {pages} {5843--5855} (\bibinfo {year} {2015})}\BibitemShut
  {NoStop}%
\bibitem [{\citenamefont {Maskowicz}\ \emph {et~al.}(2021)\citenamefont
  {Maskowicz}, \citenamefont {Sawczak}, \citenamefont {Ghosh}, \citenamefont
  {Grochowska}, \citenamefont {Jendrzejewski}, \citenamefont {Rotaru},
  \citenamefont {Garcia},\ and\ \citenamefont {Śliwiński}}]{Experiment2d-1}%
  \BibitemOpen
  \bibfield  {author} {\bibinfo {author} {\bibfnamefont {D.}~\bibnamefont
  {Maskowicz}}, \bibinfo {author} {\bibfnamefont {M.}~\bibnamefont {Sawczak}},
  \bibinfo {author} {\bibfnamefont {A.~C.}\ \bibnamefont {Ghosh}}, \bibinfo
  {author} {\bibfnamefont {K.}~\bibnamefont {Grochowska}}, \bibinfo {author}
  {\bibfnamefont {R.}~\bibnamefont {Jendrzejewski}}, \bibinfo {author}
  {\bibfnamefont {A.}~\bibnamefont {Rotaru}}, \bibinfo {author} {\bibfnamefont
  {Y.}~\bibnamefont {Garcia}}, \ and\ \bibinfo {author} {\bibfnamefont
  {G.}~\bibnamefont {Śliwiński}},\ }\bibfield  {title} {\enquote {\bibinfo
  {title} {Spin crossover and cooperativity in nanocrystalline
  [fe(pyrazine)pt(cn)4] thin films deposited by matrix-assisted laser
  evaporation},}\ }\href {\doibase
  https://doi.org/10.1016/j.apsusc.2020.148419} {\bibfield  {journal} {\bibinfo
   {journal} {Applied Surface Science}\ }\textbf {\bibinfo {volume} {541}},\
  \bibinfo {pages} {148419} (\bibinfo {year} {2021})}\BibitemShut {NoStop}%
\bibitem [{\citenamefont {Ostrovsky}\ \emph {et~al.}(2018)\citenamefont
  {Ostrovsky}, \citenamefont {Palii}, \citenamefont {Decurtins}, \citenamefont
  {Liu},\ and\ \citenamefont {Klokishner}}]{Experiment2d-2}%
  \BibitemOpen
  \bibfield  {author} {\bibinfo {author} {\bibfnamefont {S.}~\bibnamefont
  {Ostrovsky}}, \bibinfo {author} {\bibfnamefont {A.}~\bibnamefont {Palii}},
  \bibinfo {author} {\bibfnamefont {S.}~\bibnamefont {Decurtins}}, \bibinfo
  {author} {\bibfnamefont {S.-X.}\ \bibnamefont {Liu}}, \ and\ \bibinfo
  {author} {\bibfnamefont {S.}~\bibnamefont {Klokishner}},\ }\bibfield  {title}
  {\enquote {\bibinfo {title} {Microscopic approach to the problem of
  cooperative spin crossover in polynuclear cluster compounds: Application to
  tetranuclear iron(ii) square complexes},}\ }\href {\doibase
  10.1021/acs.jpcc.8b05599} {\bibfield  {journal} {\bibinfo  {journal} {The
  Journal of Physical Chemistry C}\ }\textbf {\bibinfo {volume} {122}},\
  \bibinfo {pages} {22150--22159} (\bibinfo {year} {2018})}\BibitemShut
  {NoStop}%
\bibitem [{\citenamefont {Levchenko}\ \emph {et~al.}(2014)\citenamefont
  {Levchenko}, \citenamefont {Khristov}, \citenamefont {Kuznetsova},\ and\
  \citenamefont {Shelest}}]{Experiment2d-3}%
  \BibitemOpen
  \bibfield  {author} {\bibinfo {author} {\bibfnamefont {G.}~\bibnamefont
  {Levchenko}}, \bibinfo {author} {\bibfnamefont {A.}~\bibnamefont {Khristov}},
  \bibinfo {author} {\bibfnamefont {V.}~\bibnamefont {Kuznetsova}}, \ and\
  \bibinfo {author} {\bibfnamefont {V.}~\bibnamefont {Shelest}},\ }\bibfield
  {title} {\enquote {\bibinfo {title} {Pressure and temperature induced high
  spin–low spin phase transition: Macroscopic and microscopic
  consideration},}\ }\href {\doibase
  https://doi.org/10.1016/j.jpcs.2014.04.006} {\bibfield  {journal} {\bibinfo
  {journal} {Journal of Physics and Chemistry of Solids}\ }\textbf {\bibinfo
  {volume} {75}},\ \bibinfo {pages} {966 -- 971} (\bibinfo {year}
  {2014})}\BibitemShut {NoStop}%
\bibitem [{\citenamefont {Traiche}, \citenamefont {Sy},\ and\ \citenamefont
  {Boukheddaden}(2018)}]{Spin-Crossover1d-1}%
  \BibitemOpen
  \bibfield  {author} {\bibinfo {author} {\bibfnamefont {R.}~\bibnamefont
  {Traiche}}, \bibinfo {author} {\bibfnamefont {M.}~\bibnamefont {Sy}}, \ and\
  \bibinfo {author} {\bibfnamefont {K.}~\bibnamefont {Boukheddaden}},\
  }\bibfield  {title} {\enquote {\bibinfo {title} {Elastic frustration in 1d
  spin-crossover chains: Evidence of multi-step transitions and
  self-organizations of the spin states},}\ }\href {\doibase
  https://doi.org/10.1021/acs.jpcc.7b12304} {\bibfield  {journal} {\bibinfo
  {journal} {The Journal of Physical Chemistry C}\ }\textbf {\bibinfo {volume}
  {122}},\ \bibinfo {pages} {4083--4096} (\bibinfo {year} {2018})}\BibitemShut
  {NoStop}%
\bibitem [{\citenamefont {Nicolazzi}\ \emph {et~al.}(2013)\citenamefont
  {Nicolazzi}, \citenamefont {Pavlik}, \citenamefont {Bedoui}, \citenamefont
  {Moln{\'a}r},\ and\ \citenamefont {Bousseksou}}]{Spin-Crossover1d-2}%
  \BibitemOpen
  \bibfield  {author} {\bibinfo {author} {\bibfnamefont {W.}~\bibnamefont
  {Nicolazzi}}, \bibinfo {author} {\bibfnamefont {J.}~\bibnamefont {Pavlik}},
  \bibinfo {author} {\bibfnamefont {S.}~\bibnamefont {Bedoui}}, \bibinfo
  {author} {\bibfnamefont {G.}~\bibnamefont {Moln{\'a}r}}, \ and\ \bibinfo
  {author} {\bibfnamefont {A.}~\bibnamefont {Bousseksou}},\ }\bibfield  {title}
  {\enquote {\bibinfo {title} {Elastic ising-like model for the nucleation and
  domain formation in spin crossover molecular solids},}\ }\href {\doibase
  10.1140/epjst/e2013-01911-3} {\bibfield  {journal} {\bibinfo  {journal} {The
  European Physical Journal Special Topics}\ }\textbf {\bibinfo {volume}
  {222}},\ \bibinfo {pages} {1137--1159} (\bibinfo {year} {2013})}\BibitemShut
  {NoStop}%
\bibitem [{\citenamefont {Oke}, \citenamefont {Hontinfinde},\ and\
  \citenamefont {Boukheddaden}(2015)}]{Spin-Crossover1d-3}%
  \BibitemOpen
  \bibfield  {author} {\bibinfo {author} {\bibfnamefont {T.~D.}\ \bibnamefont
  {Oke}}, \bibinfo {author} {\bibfnamefont {F.}~\bibnamefont {Hontinfinde}}, \
  and\ \bibinfo {author} {\bibfnamefont {K.}~\bibnamefont {Boukheddaden}},\
  }\bibfield  {title} {\enquote {\bibinfo {title} {Bethe lattice approach and
  relaxation dynamics study of spin-crossover materials},}\ }\href {\doibase
  https://doi.org/10.1007/s00339-015-9189-x} {\bibfield  {journal} {\bibinfo
  {journal} {Applied Physics A}\ }\textbf {\bibinfo {volume} {120}},\ \bibinfo
  {pages} {309--320} (\bibinfo {year} {2015})}\BibitemShut {NoStop}%
\bibitem [{\citenamefont {Rojas}\ \emph {et~al.}(2019)\citenamefont {Rojas},
  \citenamefont {Stre\ifmmode~\check{c}\else \v{c}\fi{}ka}, \citenamefont
  {Lyra},\ and\ \citenamefont {de~Souza}}]{PhysRevE.99.042117}%
  \BibitemOpen
  \bibfield  {author} {\bibinfo {author} {\bibfnamefont {O.}~\bibnamefont
  {Rojas}}, \bibinfo {author} {\bibfnamefont {J.}~\bibnamefont
  {Stre\ifmmode~\check{c}\else \v{c}\fi{}ka}}, \bibinfo {author} {\bibfnamefont
  {M.~L.}\ \bibnamefont {Lyra}}, \ and\ \bibinfo {author} {\bibfnamefont
  {S.~M.}\ \bibnamefont {de~Souza}},\ }\bibfield  {title} {\enquote {\bibinfo
  {title} {Universality and quasicritical exponents of one-dimensional models
  displaying a quasitransition at finite temperatures},}\ }\href {\doibase
  10.1103/PhysRevE.99.042117} {\bibfield  {journal} {\bibinfo  {journal} {Phys.
  Rev. E}\ }\textbf {\bibinfo {volume} {99}},\ \bibinfo {pages} {042117}
  (\bibinfo {year} {2019})}\BibitemShut {NoStop}%
\bibitem [{\citenamefont {Hutak}\ \emph {et~al.}(2021)\citenamefont {Hutak},
  \citenamefont {Krokhmalskii}, \citenamefont {Rojas}, \citenamefont {{Martins
  de Souza}},\ and\ \citenamefont {Derzhko}}]{HUTAK2021127020}%
  \BibitemOpen
  \bibfield  {author} {\bibinfo {author} {\bibfnamefont {T.}~\bibnamefont
  {Hutak}}, \bibinfo {author} {\bibfnamefont {T.}~\bibnamefont {Krokhmalskii}},
  \bibinfo {author} {\bibfnamefont {O.}~\bibnamefont {Rojas}}, \bibinfo
  {author} {\bibfnamefont {S.}~\bibnamefont {{Martins de Souza}}}, \ and\
  \bibinfo {author} {\bibfnamefont {O.}~\bibnamefont {Derzhko}},\ }\bibfield
  {title} {\enquote {\bibinfo {title} {Low-temperature thermodynamics of the
  two-leg ladder ising model with trimer rungs: A mystery explained},}\ }\href
  {\doibase https://doi.org/10.1016/j.physleta.2020.127020} {\bibfield
  {journal} {\bibinfo  {journal} {Physics Letters A}\ }\textbf {\bibinfo
  {volume} {387}},\ \bibinfo {pages} {127020} (\bibinfo {year}
  {2021})}\BibitemShut {NoStop}%
\bibitem [{\citenamefont {Gudyma}\ and\ \citenamefont
  {Gudyma}(2021)}]{gudyma20211d}%
  \BibitemOpen
  \bibfield  {author} {\bibinfo {author} {\bibfnamefont {A.}~\bibnamefont
  {Gudyma}}\ and\ \bibinfo {author} {\bibfnamefont {{\relax Iu}.}~\bibnamefont
  {Gudyma}},\ }\href@noop {} {\enquote {\bibinfo {title} {1d spin-crossover
  molecular chain with degenerate states},}\ } (\bibinfo {year} {2021}),\
  \Eprint {http://arxiv.org/abs/2102.13627} {arXiv:2102.13627
  [cond-mat.stat-mech]} \BibitemShut {NoStop}%
\bibitem [{\citenamefont {Zagrebnov}\ and\ \citenamefont
  {Fedyanin}(1972)}]{zagrebnov1972spin}%
  \BibitemOpen
  \bibfield  {author} {\bibinfo {author} {\bibfnamefont {V.~A.}\ \bibnamefont
  {Zagrebnov}}\ and\ \bibinfo {author} {\bibfnamefont {B.~K.}\ \bibnamefont
  {Fedyanin}},\ }\bibfield  {title} {\enquote {\bibinfo {title} {Spin-phonon
  interaction in the ising model},}\ }\href {\doibase 10.1007/BF01035771}
  {\bibfield  {journal} {\bibinfo  {journal} {Theoretical and Mathematical
  Physics}\ }\textbf {\bibinfo {volume} {10}},\ \bibinfo {pages} {84--93}
  (\bibinfo {year} {1972})}\BibitemShut {NoStop}%
\bibitem [{\citenamefont {Salinas}(1973)}]{Salinas_1973}%
  \BibitemOpen
  \bibfield  {author} {\bibinfo {author} {\bibfnamefont {S.~R.}\ \bibnamefont
  {Salinas}},\ }\bibfield  {title} {\enquote {\bibinfo {title} {On the
  one-dimensional compressible ising model},}\ }\href {\doibase
  10.1088/0305-4470/6/10/011} {\bibfield  {journal} {\bibinfo  {journal}
  {Journal of Physics A: Mathematical, Nuclear and General}\ }\textbf {\bibinfo
  {volume} {6}},\ \bibinfo {pages} {1527--1533} (\bibinfo {year}
  {1973})}\BibitemShut {NoStop}%
\bibitem [{\citenamefont {Henriques}\ and\ \citenamefont
  {Salinas}(1987)}]{Henriques_1987}%
  \BibitemOpen
  \bibfield  {author} {\bibinfo {author} {\bibfnamefont {V.~B.}\ \bibnamefont
  {Henriques}}\ and\ \bibinfo {author} {\bibfnamefont {S.~R.}\ \bibnamefont
  {Salinas}},\ }\bibfield  {title} {\enquote {\bibinfo {title} {Effective spin
  hamiltonians for compressible ising models},}\ }\href {\doibase
  10.1088/0022-3719/20/16/014} {\bibfield  {journal} {\bibinfo  {journal}
  {Journal of Physics C: Solid State Physics}\ }\textbf {\bibinfo {volume}
  {20}},\ \bibinfo {pages} {2415--2429} (\bibinfo {year} {1987})}\BibitemShut
  {NoStop}%
\bibitem [{\citenamefont {Marshall}, \citenamefont {Chakraborty},\ and\
  \citenamefont {Nagel}(2006)}]{Numeric-1}%
  \BibitemOpen
  \bibfield  {author} {\bibinfo {author} {\bibfnamefont {A.~H.}\ \bibnamefont
  {Marshall}}, \bibinfo {author} {\bibfnamefont {B.}~\bibnamefont
  {Chakraborty}}, \ and\ \bibinfo {author} {\bibfnamefont {S.}~\bibnamefont
  {Nagel}},\ }\bibfield  {title} {\enquote {\bibinfo {title} {Numerical studies
  of the compressible ising spin glass},}\ }\href {\doibase
  10.1209/epl/i2005-10564-5} {\bibfield  {journal} {\bibinfo  {journal}
  {Europhysics Letters ({EPL})}\ }\textbf {\bibinfo {volume} {74}},\ \bibinfo
  {pages} {699--705} (\bibinfo {year} {2006})}\BibitemShut {NoStop}%
\bibitem [{\citenamefont {Balcerzak}, \citenamefont {Szałowski},\ and\
  \citenamefont {Jaščur}(2020)}]{Numeric-2}%
  \BibitemOpen
  \bibfield  {author} {\bibinfo {author} {\bibfnamefont {T.}~\bibnamefont
  {Balcerzak}}, \bibinfo {author} {\bibfnamefont {K.}~\bibnamefont
  {Szałowski}}, \ and\ \bibinfo {author} {\bibfnamefont {M.}~\bibnamefont
  {Jaščur}},\ }\bibfield  {title} {\enquote {\bibinfo {title} {Thermodynamic
  properties of the one-dimensional ising model with magnetoelastic
  interaction},}\ }\href {\doibase https://doi.org/10.1016/j.jmmm.2020.166825}
  {\bibfield  {journal} {\bibinfo  {journal} {Journal of Magnetism and Magnetic
  Materials}\ }\textbf {\bibinfo {volume} {507}},\ \bibinfo {pages} {166825}
  (\bibinfo {year} {2020})}\BibitemShut {NoStop}%
\bibitem [{\citenamefont {Nakada}\ \emph {et~al.}(2012)\citenamefont {Nakada},
  \citenamefont {Mori}, \citenamefont {Miyashita}, \citenamefont {Nishino},
  \citenamefont {Todo}, \citenamefont {Nicolazzi},\ and\ \citenamefont
  {Rikvold}}]{Numeric-3}%
  \BibitemOpen
  \bibfield  {author} {\bibinfo {author} {\bibfnamefont {T.}~\bibnamefont
  {Nakada}}, \bibinfo {author} {\bibfnamefont {T.}~\bibnamefont {Mori}},
  \bibinfo {author} {\bibfnamefont {S.}~\bibnamefont {Miyashita}}, \bibinfo
  {author} {\bibfnamefont {M.}~\bibnamefont {Nishino}}, \bibinfo {author}
  {\bibfnamefont {S.}~\bibnamefont {Todo}}, \bibinfo {author} {\bibfnamefont
  {W.}~\bibnamefont {Nicolazzi}}, \ and\ \bibinfo {author} {\bibfnamefont
  {P.~A.}\ \bibnamefont {Rikvold}},\ }\bibfield  {title} {\enquote {\bibinfo
  {title} {Critical temperature and correlation length of an elastic
  interaction model for spin-crossover materials},}\ }\href {\doibase
  10.1103/PhysRevB.85.054408} {\bibfield  {journal} {\bibinfo  {journal} {Phys.
  Rev. B}\ }\textbf {\bibinfo {volume} {85}},\ \bibinfo {pages} {054408}
  (\bibinfo {year} {2012})}\BibitemShut {NoStop}%
\bibitem [{\citenamefont {Ye}, \citenamefont {Sun},\ and\ \citenamefont
  {Jiang}(2015)}]{Numeric-4}%
  \BibitemOpen
  \bibfield  {author} {\bibinfo {author} {\bibfnamefont {H.-Z.}\ \bibnamefont
  {Ye}}, \bibinfo {author} {\bibfnamefont {C.}~\bibnamefont {Sun}}, \ and\
  \bibinfo {author} {\bibfnamefont {H.}~\bibnamefont {Jiang}},\ }\bibfield
  {title} {\enquote {\bibinfo {title} {Monte-carlo simulations of
  spin-crossover phenomena based on a vibronic ising-like model with realistic
  parameters},}\ }\href {\doibase 10.1039/C4CP05562D} {\bibfield  {journal}
  {\bibinfo  {journal} {Phys. Chem. Chem. Phys.}\ }\textbf {\bibinfo {volume}
  {17}},\ \bibinfo {pages} {6801--6808} (\bibinfo {year} {2015})}\BibitemShut
  {NoStop}%
\bibitem [{\citenamefont {Banerjee}, \citenamefont {Kumar},\ and\ \citenamefont
  {Saha-Dasgupta}(2014)}]{Numeric-5}%
  \BibitemOpen
  \bibfield  {author} {\bibinfo {author} {\bibfnamefont {H.}~\bibnamefont
  {Banerjee}}, \bibinfo {author} {\bibfnamefont {M.}~\bibnamefont {Kumar}}, \
  and\ \bibinfo {author} {\bibfnamefont {T.}~\bibnamefont {Saha-Dasgupta}},\
  }\bibfield  {title} {\enquote {\bibinfo {title} {Cooperativity in
  spin-crossover transition in metalorganic complexes: Interplay of magnetic
  and elastic interactions},}\ }\href {\doibase 10.1103/PhysRevB.90.174433}
  {\bibfield  {journal} {\bibinfo  {journal} {Phys. Rev. B}\ }\textbf {\bibinfo
  {volume} {90}},\ \bibinfo {pages} {174433} (\bibinfo {year}
  {2014})}\BibitemShut {NoStop}%
\bibitem [{\citenamefont {Apetrei}, \citenamefont {Boukheddaden},\ and\
  \citenamefont {Stancu}(2013)}]{Numeric-6}%
  \BibitemOpen
  \bibfield  {author} {\bibinfo {author} {\bibfnamefont {A.~M.}\ \bibnamefont
  {Apetrei}}, \bibinfo {author} {\bibfnamefont {K.}~\bibnamefont
  {Boukheddaden}}, \ and\ \bibinfo {author} {\bibfnamefont {A.}~\bibnamefont
  {Stancu}},\ }\bibfield  {title} {\enquote {\bibinfo {title} {Dynamic phase
  transitions in the one-dimensional spin-phonon coupling model},}\ }\href
  {\doibase 10.1103/PhysRevB.87.014302} {\bibfield  {journal} {\bibinfo
  {journal} {Phys. Rev. B}\ }\textbf {\bibinfo {volume} {87}},\ \bibinfo
  {pages} {014302} (\bibinfo {year} {2013})}\BibitemShut {NoStop}%
\bibitem [{\citenamefont {Gaspar}\ \emph {et~al.}(2018)\citenamefont {Gaspar},
  \citenamefont {Molnár}, \citenamefont {Rotaru},\ and\ \citenamefont
  {Shepherd}}]{Pressure-1}%
  \BibitemOpen
  \bibfield  {author} {\bibinfo {author} {\bibfnamefont {A.~B.}\ \bibnamefont
  {Gaspar}}, \bibinfo {author} {\bibfnamefont {G.}~\bibnamefont {Molnár}},
  \bibinfo {author} {\bibfnamefont {A.}~\bibnamefont {Rotaru}}, \ and\ \bibinfo
  {author} {\bibfnamefont {H.~J.}\ \bibnamefont {Shepherd}},\ }\bibfield
  {title} {\enquote {\bibinfo {title} {Pressure effect investigations on
  spin-crossover coordination compounds},}\ }\href {\doibase
  https://doi.org/10.1016/j.crci.2018.07.010} {\bibfield  {journal} {\bibinfo
  {journal} {Comptes Rendus Chimie}\ }\textbf {\bibinfo {volume} {21}},\
  \bibinfo {pages} {1095 -- 1120} (\bibinfo {year} {2018})}\BibitemShut
  {NoStop}%
\bibitem [{\citenamefont {Gudyma}, \citenamefont {Ivashko},\ and\ \citenamefont
  {Linares}(2014)}]{gudyma2014diffusionless}%
  \BibitemOpen
  \bibfield  {author} {\bibinfo {author} {\bibfnamefont {{\relax
  Iu}.}~\bibnamefont {Gudyma}}, \bibinfo {author} {\bibfnamefont
  {V.}~\bibnamefont {Ivashko}}, \ and\ \bibinfo {author} {\bibfnamefont
  {J.}~\bibnamefont {Linares}},\ }\bibfield  {title} {\enquote {\bibinfo
  {title} {Diffusionless phase transition with two order parameters in
  spin-crossover solids},}\ }\href {\doibase https://doi.org/10.1063/1.4901243}
  {\bibfield  {journal} {\bibinfo  {journal} {Journal of Applied Physics}\
  }\textbf {\bibinfo {volume} {116}},\ \bibinfo {pages} {173509} (\bibinfo
  {year} {2014})}\BibitemShut {NoStop}%
\bibitem [{\citenamefont {Gudyma}, \citenamefont {Maksymov},\ and\
  \citenamefont {Ivashko}(2014)}]{gudyma2014study}%
  \BibitemOpen
  \bibfield  {author} {\bibinfo {author} {\bibfnamefont {{\relax Iu}.~V.}\
  \bibnamefont {Gudyma}}, \bibinfo {author} {\bibfnamefont {A.~I.}\
  \bibnamefont {Maksymov}}, \ and\ \bibinfo {author} {\bibfnamefont {V.~V.}\
  \bibnamefont {Ivashko}},\ }\bibfield  {title} {\enquote {\bibinfo {title}
  {Study of pressure influence on thermal transition in spin-crossover
  nanomaterials},}\ }\href {\doibase https://doi.org/10.1186/1556-276X-9-691}
  {\bibfield  {journal} {\bibinfo  {journal} {Nanoscale Research Letters}\
  }\textbf {\bibinfo {volume} {9}},\ \bibinfo {pages} {1--6} (\bibinfo {year}
  {2014})}\BibitemShut {NoStop}%
\bibitem [{\citenamefont {Gudyma}\ and\ \citenamefont
  {Ivashko}(2016)}]{gudyma2016spin}%
  \BibitemOpen
  \bibfield  {author} {\bibinfo {author} {\bibfnamefont {{\relax Iu}.~V.}\
  \bibnamefont {Gudyma}}\ and\ \bibinfo {author} {\bibfnamefont {V.~V.}\
  \bibnamefont {Ivashko}},\ }\bibfield  {title} {\enquote {\bibinfo {title}
  {Spin-crossover molecular solids beyond rigid crystal approximation},}\
  }\href {\doibase https://doi.org/10.1186/s11671-016-1398-5} {\bibfield
  {journal} {\bibinfo  {journal} {Nanoscale Research Letters}\ }\textbf
  {\bibinfo {volume} {11}},\ \bibinfo {pages} {196} (\bibinfo {year}
  {2016})}\BibitemShut {NoStop}%
\bibitem [{\citenamefont {Nicolazzi}, \citenamefont {Pillet},\ and\
  \citenamefont {Lecomte}(2008)}]{Elastic-potential-1}%
  \BibitemOpen
  \bibfield  {author} {\bibinfo {author} {\bibfnamefont {W.}~\bibnamefont
  {Nicolazzi}}, \bibinfo {author} {\bibfnamefont {S.}~\bibnamefont {Pillet}}, \
  and\ \bibinfo {author} {\bibfnamefont {C.}~\bibnamefont {Lecomte}},\
  }\bibfield  {title} {\enquote {\bibinfo {title} {Two-variable anharmonic
  model for spin-crossover solids: A like-spin domains interpretation},}\
  }\href {\doibase 10.1103/PhysRevB.78.174401} {\bibfield  {journal} {\bibinfo
  {journal} {Phys. Rev. B}\ }\textbf {\bibinfo {volume} {78}},\ \bibinfo
  {pages} {174401} (\bibinfo {year} {2008})}\BibitemShut {NoStop}%
\bibitem [{\citenamefont {Legrand}\ \emph {et~al.}(2007)\citenamefont
  {Legrand}, \citenamefont {Pillet}, \citenamefont {Carbonera}, \citenamefont
  {Souhassou}, \citenamefont {L{\'e}tard}, \citenamefont {Guionneau},\ and\
  \citenamefont {Lecomte}}]{Experiment1d-Lecomte}%
  \BibitemOpen
  \bibfield  {author} {\bibinfo {author} {\bibfnamefont {V.}~\bibnamefont
  {Legrand}}, \bibinfo {author} {\bibfnamefont {S.}~\bibnamefont {Pillet}},
  \bibinfo {author} {\bibfnamefont {C.}~\bibnamefont {Carbonera}}, \bibinfo
  {author} {\bibfnamefont {M.}~\bibnamefont {Souhassou}}, \bibinfo {author}
  {\bibfnamefont {J.-F.}\ \bibnamefont {L{\'e}tard}}, \bibinfo {author}
  {\bibfnamefont {P.}~\bibnamefont {Guionneau}}, \ and\ \bibinfo {author}
  {\bibfnamefont {C.}~\bibnamefont {Lecomte}},\ }\bibfield  {title} {\enquote
  {\bibinfo {title} {Optical, magnetic and structural properties of the
  spin-crossover complex [fe (btr) 2 (ncs) 2]{\textperiodcentered} h 2 o in the
  light-induced and thermally quenched metastable states},}\ }\href {\doibase
  10.1002/ejic.200700872} {\bibfield  {journal} {\bibinfo  {journal} {European
  Journal of Inorganic Chemistry}\ }\textbf {\bibinfo {volume} {36}},\ \bibinfo
  {pages} {5693--5706} (\bibinfo {year} {2007})}\BibitemShut {NoStop}%
\bibitem [{\citenamefont {Jung}\ \emph {et~al.}(1996)\citenamefont {Jung},
  \citenamefont {Bruchh{\"a}user}, \citenamefont {Feile}, \citenamefont
  {Spiering},\ and\ \citenamefont {G{\"u}tlich}}]{Experiment1d-Gutlich}%
  \BibitemOpen
  \bibfield  {author} {\bibinfo {author} {\bibfnamefont {J.}~\bibnamefont
  {Jung}}, \bibinfo {author} {\bibfnamefont {F.}~\bibnamefont
  {Bruchh{\"a}user}}, \bibinfo {author} {\bibfnamefont {R.}~\bibnamefont
  {Feile}}, \bibinfo {author} {\bibfnamefont {H.}~\bibnamefont {Spiering}}, \
  and\ \bibinfo {author} {\bibfnamefont {P.}~\bibnamefont {G{\"u}tlich}},\
  }\bibfield  {title} {\enquote {\bibinfo {title} {The cooperative spin
  transition in [fe x zn 1- x (ptz) 6](bf 4) 2: I. elastic properties—an
  oriented sample rotation study by brillouin spectroscopy},}\ }\href {\doibase
  10.1007/s002570050156} {\bibfield  {journal} {\bibinfo  {journal}
  {Zeitschrift f{\"u}r Physik B Condensed Matter}\ }\textbf {\bibinfo {volume}
  {100}},\ \bibinfo {pages} {517--522} (\bibinfo {year} {1996})}\BibitemShut
  {NoStop}%
\bibitem [{\citenamefont {Bousseksou}\ \emph {et~al.}(1992)\citenamefont
  {Bousseksou}, \citenamefont {Nasser}, \citenamefont {Linares}, \citenamefont
  {Boukheddaden},\ and\ \citenamefont {Varret}}]{bousseksou1992ising}%
  \BibitemOpen
  \bibfield  {author} {\bibinfo {author} {\bibfnamefont {A.}~\bibnamefont
  {Bousseksou}}, \bibinfo {author} {\bibfnamefont {J.}~\bibnamefont {Nasser}},
  \bibinfo {author} {\bibfnamefont {J.}~\bibnamefont {Linares}}, \bibinfo
  {author} {\bibfnamefont {K.}~\bibnamefont {Boukheddaden}}, \ and\ \bibinfo
  {author} {\bibfnamefont {F.}~\bibnamefont {Varret}},\ }\bibfield  {title}
  {\enquote {\bibinfo {title} {Ising-like model for the two-step
  spin-crossover},}\ }\href@noop {} {\bibfield  {journal} {\bibinfo  {journal}
  {Journal De Physique I}\ }\textbf {\bibinfo {volume} {2}},\ \bibinfo {pages}
  {1381--1403} (\bibinfo {year} {1992})}\BibitemShut {NoStop}%
\bibitem [{\citenamefont {Boukheddaden}\ \emph {et~al.}(2000)\citenamefont
  {Boukheddaden}, \citenamefont {Linares}, \citenamefont {Spiering},\ and\
  \citenamefont {Varret}}]{boukheddaden2000one}%
  \BibitemOpen
  \bibfield  {author} {\bibinfo {author} {\bibfnamefont {K.}~\bibnamefont
  {Boukheddaden}}, \bibinfo {author} {\bibfnamefont {J.}~\bibnamefont
  {Linares}}, \bibinfo {author} {\bibfnamefont {H.}~\bibnamefont {Spiering}}, \
  and\ \bibinfo {author} {\bibfnamefont {F.}~\bibnamefont {Varret}},\
  }\bibfield  {title} {\enquote {\bibinfo {title} {One-dimensional ising-like
  systems: an analytical investigation of the static and dynamic properties,
  applied to spin-crossover relaxation},}\ }\href@noop {} {\bibfield  {journal}
  {\bibinfo  {journal} {The European Physical Journal B-Condensed Matter and
  Complex Systems}\ }\textbf {\bibinfo {volume} {15}},\ \bibinfo {pages}
  {317--326} (\bibinfo {year} {2000})}\BibitemShut {NoStop}%
\bibitem [{\citenamefont {Nasser}, \citenamefont {Boukheddaden},\ and\
  \citenamefont {Linares}(2004)}]{Linares2004}%
  \BibitemOpen
  \bibfield  {author} {\bibinfo {author} {\bibfnamefont {J.}~\bibnamefont
  {Nasser}}, \bibinfo {author} {\bibfnamefont {K.}~\bibnamefont
  {Boukheddaden}}, \ and\ \bibinfo {author} {\bibfnamefont {J.}~\bibnamefont
  {Linares}},\ }\bibfield  {title} {\enquote {\bibinfo {title} {Two-step spin
  conversion and other effects in the atom-phonon coupling model},}\ }\href
  {\doibase 10.1140/epjb/e2004-00184-y} {\bibfield  {journal} {\bibinfo
  {journal} {The European Physical Journal B-Condensed Matter and Complex
  Systems}\ }\textbf {\bibinfo {volume} {39}},\ \bibinfo {pages} {219--227}
  (\bibinfo {year} {2004})}\BibitemShut {NoStop}%
\bibitem [{\citenamefont {Boukheddaden}, \citenamefont {Miyashita},\ and\
  \citenamefont {Nishino}(2007)}]{Boukheddaden2007}%
  \BibitemOpen
  \bibfield  {author} {\bibinfo {author} {\bibfnamefont {K.}~\bibnamefont
  {Boukheddaden}}, \bibinfo {author} {\bibfnamefont {S.}~\bibnamefont
  {Miyashita}}, \ and\ \bibinfo {author} {\bibfnamefont {M.}~\bibnamefont
  {Nishino}},\ }\bibfield  {title} {\enquote {\bibinfo {title} {Elastic
  interaction among transition metals in one-dimensional spin-crossover
  solids},}\ }\href {\doibase 10.1103/PhysRevB.75.094112} {\bibfield  {journal}
  {\bibinfo  {journal} {Phys. Rev. B}\ }\textbf {\bibinfo {volume} {75}},\
  \bibinfo {pages} {094112} (\bibinfo {year} {2007})}\BibitemShut {NoStop}%
\bibitem [{\citenamefont {Bellucci}\ and\ \citenamefont
  {Ohanyan}(2013)}]{bellucci2013correlation}%
  \BibitemOpen
  \bibfield  {author} {\bibinfo {author} {\bibfnamefont {S.}~\bibnamefont
  {Bellucci}}\ and\ \bibinfo {author} {\bibfnamefont {V.}~\bibnamefont
  {Ohanyan}},\ }\bibfield  {title} {\enquote {\bibinfo {title} {Correlation
  functions in one-dimensional spin lattices with {Ising} and heisenberg
  bonds},}\ }\href {\doibase https://doi.org/10.1140/epjb/e2013-40336-4}
  {\bibfield  {journal} {\bibinfo  {journal} {The European Physical Journal B}\
  }\textbf {\bibinfo {volume} {86}},\ \bibinfo {pages} {446} (\bibinfo {year}
  {2013})}\BibitemShut {NoStop}%
\end{thebibliography}%

\end{document}